\documentclass[aps,prstab,preprint,groupedaddress,showpacs]{revtex4}
\usepackage[dvips]{graphicx}
\begin{document}

%\preprint{}

\title{Chaotic orbits in thermal-equilibrium beams: existence and dynamical implications}


\author{Courtlandt L. Bohn$^{1,2}$ and Ioannis V. Sideris$^1$}
%\email[clbohn@fnal.gov]{Your e-mail address}
%\thanks{}
\affiliation{$^1$Northern Illinois University, DeKalb, IL 60115
\\
$^2$Fermilab, Batavia, IL 60115}


\date{\today}

\begin{abstract}
Phase mixing of chaotic orbits exponentially distributes these orbits through their accessible phase space.
This phenomenon, commonly called ``chaotic mixing'', stands in marked contrast to phase mixing of regular orbits which proceeds as a power law in time.
It is operationally irreversible; hence, its associated e-folding time scale sets a condition on any process envisioned for emittance compensation.
A key question is whether beams can support chaotic orbits, and if so, under what conditions?
We numerically investigate the parameter space of three-dimensional thermal-equilibrium beams with space charge, confined by linear external focusing forces, to determine whether the associated potentials support chaotic orbits.
We find that a large subset of the parameter space does support chaos and, in turn, chaotic mixing.
Details and implications are enumerated.
\end{abstract}

%\keywords{}

\maketitle

\section{Introduction\label{sec:intro}}
Rapid, inherently irreversible dynamics is a practical concern in producing high-brightness charged-particle beams.
Time scales of irreversible processes place constraints on methods for compensating against degradation of beam quality caused by, for example, space charge.
This is a very important practical matter because compensation must be fast compared to these processes, and this affects the choice and configuration of the associated hardware.

A beam bunch with space charge comprises an $\it{N}$-body system with typically 3$\it{N}$ degrees of freedom.
Upon coarse-graining, i.e., ``smoothing'' the system to remove granularity, the collective space-charge force remains.
One might conjecture that this force, when nonlinear, may support chaotic orbits.
One example is the University of Maryland five-beamlet experiment that shows presumably irreversible dissipation of the beamlets after a few space-charge-depressed betatron periods~\cite{reiser94}.
Simulations of the experiment reveal a substantial fraction of globally chaotic orbits~\cite{bohn02}, and phase mixing of these orbits thereby presents itself as a contributing evolutionary mechanism.
This example pertains to a strongly time-dependent nonequilibrium system, yet one might conjecture that nonlinear space-charge forces in a static system could support chaotic orbits as well.
We shall explore this conjecture.

An initially localized clump of chaotic orbits will, via phase mixing, grow
exponentially and eventually reach an invariant distribution.
This is ``chaotic mixing''~\cite{kandrup94,merritt96}.
Strictly speaking, the process is reversible in that it is collisionless and its dynamics is included in, e.g., Vlasov's equation.
Nonetheless, when the invariant distribution spans a global region of the system's phase space, chaotic mixing is a legitimate relaxation mechanism in that it drastically smears correlations.
Moreover, from a practical perspective, the process is strictly irreversible because infinitesimal fine-tuning is needed to reassemble the initial conditions.
It is also distinctly different from phase mixing of regular orbits, i.e., linear Landau damping~\cite{sagan94}, a process that winds an initially localized clump into a filament over a comparatively narrow region of phase space.
Whereas chaotic mixing proceeds exponentially over a well-defined time scale and can cause global, macroscopic changes in the system, phase mixing of regular orbits carries a power-law time dependence, proceeds on a time scale depending on the distribution of orbital frequencies across the clump, and acts only over a portion of the phase space.
Accordingly, ascertaining conditions for, and time scales of, chaotic mixing in beams is an undertaking of practical importance.

In this paper we consider a family of thermal-equilibrium (TE) configurations of beam bunches with space charge, i.e., nonneutral plasmas, confined by linear external forces~\cite{davidson99,brown95,reiser93}.
For simplicity, we treat the dynamics in a reference frame that comoves with the bunch and has its origin affixed to the bunch centroid.
Particle motion in this reference frame is taken to be nonrelativistic; transforming from the bunch frame to the laboratory frame is straightforward~\cite{reiser2}.
In the laboratory frame the space-charge force decreases inversely with the square of the beam energy.
For the transverse component, this arises from the partial cancellation between the self-magnetic and self-electrostatic forces; while for the longitudinal component, it is due to Lorentz contraction~\cite{chao93}.
Nonetheless, there are many situations involving high-brightness beams wherein space charge is important.
Contemporary examples include low-to-medium-energy hadron accelerators such as those that drive spallation-neutron sources or serve as boosters for high-energy machines, heavy-ion accelerators, and low-energy electron accelerators such as photoinjectors~\cite{lee96}.

Thermal-equilibrium beams are of practical interest in connection with, e.g., high-current radiofrequency linear accelerators.
While conventional designs of such machines lead to bunches that are out of equilibrium, a design strategy that keeps the beam at or near thermal equilibrium has been formulated~\cite{reiser95}.
The principal motivation for this alternative strategy is to circumvent equipartitioning processes that cause emittance growth and halo formation.

Because a TE configuration is a maximum-entropy configuration, is static, and is manifestly stable~\cite{davidson98}, one might expect its intrinsic dynamics to be entirely benign.
The expectation is questionable.
The density distribution of such a configuration is uniform in its interior and falls to zero over a distance commensurate to the Debye length.
Thus, large-amplitude orbits will explore this ``Debye tail,'' during which time they experience a nonlinear force.
The question we seek to answer is whether the nonlinear force in the Debye tail can cause a significant number of orbits to be chaotic.
The answer is unequivocally ``no'' for spherically symmetric or infinitely long cylindrically symmetric configurations because their potentials are integrable and thereby support only regular orbits.
However, breaking the symmetry {\it can} generate chaotic orbits, as will become apparent in the analysis to follow.

Our study involves a comprehensive suite of numerical experiments concerning orbital dynamics in smooth (coarse-grained) TE configurations.
We establish a quantitative measure of chaos in orbits and use this measure to distinguish between regular and chaotic orbits.
We then evolve initially localized clumps of particles in the smooth potentials.
The experiments are fast if the potentials are analytic, but they are much slower if the potentials must first be tabulated numerically over a grid.
As part of the preliminaries, Sec.~\ref{sec:mixingtime} presents a semianalytic theory for estimating the time scale for chaotic mixing.
In general the TE configurations, specified in Sec.~\ref{sec:TE}, must be found numerically.
Section~\ref{sec:appx} presents a means for rapidly constructing approximate, semianalytic models of their potentials.
With these models we are able to survey the parameter space and obtain a zeroth-order assessment of the prevalence and degree of chaos; this is done in Sec.~\ref{sec:survey}.
Section~\ref{sec:exact} concerns examples for which the potential is accurately determined via a numerical solution of Poisson's equation on a grid.
For these examples the experiments of Sec.~\ref{sec:survey} are repeated, and the results are compared to those derived from the semianalytic approximation.
Section~\ref{sec:summary} summarizes the findings, discusses their implications while providing a comparison with the theory of Sec.~\ref{sec:mixingtime}, and presents a path for follow-on work.

\section{Estimated Time Scale for Chaotic Mixing
\label{sec:mixingtime}}
Before embarking on numerical studies, it is wise to ascertain whether chaotic mixing can indeed proceed rapidly.
One can construct an analytic tool to estimate the chaotic-mixing rate, although its application involves the tacit assumption, or initial knowledge, that chaotic orbits are present.
In this section we sketch the methodology leading to analytic predictions.
Additional details are available elsewhere~\cite{bohn00,kandrup01}.

The past few years have seen development of a geometric method proposed by M. Pettini to quantify chaotic instability in Hamiltonian systems with many degrees of freedom.
The central idea is to describe the dynamics in terms of average curvature properties of the manifold in which the particle orbits are geodesics.
The method hinges on the following assumptions and approximations; they are discussed thoroughly in Ref.~\cite{casetti96}: (1) a generic geodesic is chaotic; (2) the manifold's effective curvature is locally deformed but otherwise constant; (3) the effective curvature reflects a gaussian stochastic process; and (4) long-time-averaged properties of the curvature are calculable as phase-space averages over an invariant measure, specifically, the microcanonical ensemble.
The gaussian process is the zeroth-order term in a cumulant expansion of the actual stochastic process; assumption (3) is that the zeroth-order term suffices.
The end result relates chaotic instability to the geometric properties of the manifold defined by the long-time-averaged orbits.
In short, the theory is based on (often questionable) assumptions that chaos exists and is characterized by ergodicity and a microcanonical ensemble, and it treats chaotic orbits as arising from a parametric instability that can be modeled by a stochastic-oscillator equation.
It has recently been adapted for application to low-dimensional, autonomous (time-independent) Hamiltonian systems and, in tests against a wide variety of such systems, it was found commonly to yield estimates of mixing rates that are good to within a factor $\sim 2$~\cite{kandrup01}.

Action principles in classical mechanics are tantamount to extremals of ``arc lengths''; thus, one can infer a metric tensor from an action principle~\cite{goldstein50}.
The metric tensor manifests all of the properties of the manifold over which the system evolves, with these properties being calculable following standard methods of differential geometry.
Of special interest is the divergence of two initially nearby $3N$-dimensional geodesics ${\bf q}$ and ${\bf q}+\delta{\bf q}$ as governed by the equation of geodesic deviation:
\begin{equation}
\frac{D^2\delta q^\alpha}{ds^2}
+{R^\alpha}_{\beta\gamma\delta}\frac{dq^\beta}{ds}
\delta q^\gamma\frac{dq^\delta}{ds}~=~0,
\label{eq:geodev}
\end{equation}
in which $D/ds$ denotes covariant differentiation with respect to the ``proper time'' $s$, ${R^\alpha}_{\beta\gamma\delta}$ is the Riemann tensor derivable from the metric tensor, and summation over repeated indices is implied with each index spanning the $3N$ degrees of freedom.
Equation~(\ref{eq:geodev}) is the basis for determining the mixing rate $\chi$ as a measure of the system's largest Lyapunov exponent, a quantity that reflects the long-time behavior of the separation vector:
\begin{equation}
\chi~=~\lim_{t\rightarrow\infty}\frac{1}{t}
\ln\frac{|\delta{\bf q}(t)|}{|\delta{\bf q}(0)|}.
\label{eq:lyapunov}
\end{equation}

Any number of action principles, and therefore any number of metric tensors, can be selected to proceed further.
Eisenhart's metric~\cite{eisenhart29}, which is consistent with Hamilton's least-action principle, is probably the most convenient choice. 
It offers easy calculation of the Riemann tensor, and it avoids spurious results traceable to the singular boundary of the perhaps better-known Jacobi metric that is derivable from Maupertius' least-action principle~\cite{szczesny99}.
Eisenhart's metric operates over an enlarged configuration space-time manifold in which the geodesics are parameterized by the real time $t$, i.e., $ds^2=dt^2=-2V({\bf q})(dq^0)^2+\delta_{ij}dq^idq^j+2dq^0dq^{3N+1}$, in which $V({\bf q})$ is the potential energy per unit mass (hereafter called the ``potential''); $\delta_{ij}$ (with the indices $i,j$ running from $1$ to $3N$) is the unit tensor corresponding (without loss of generality) to a cartesian spatial coordinate system, $q^0=t$; $q^{3N+1}=t/2-\int_0^t dt^\prime L({\bf q},\dot{\bf q})$; and $L$ is the Lagrangian.
The resulting geodesic equations for the spatial coordinates $q^i$ are Newton's equations of motion, so the particle trajectories correspond to a canonical projection of the Eisenhart geodesics onto the configuration space-time manifold.
A convenient byproduct of the Eisenhart metric is that the only nonzero components of the Riemann tensor are $R_{0i0j}=\partial_i\partial_j V$, in which $\partial_i=\partial/\partial q^i$.

Using the aforementioned assumptions and approximations, Pettini and others~\cite{casetti96,cipriani98} derive an expression for $\chi$ in terms of the curvature and its standard deviation averaged over the microcanonical ensemble.
The idea is that, as $t\rightarrow\infty$, chaotic orbits of total energy $E$ mix through the configuration space toward an invariant measure, taken per assumption (4) to be the microcanonical ensemble $\delta (H-E)$, over which time averages become equivalent to phase-space averages.
Specifically, for an arbitrary function $A({\bf q})$, the averaging process is
\begin{equation}
\langle A\rangle~\equiv~\lim_{t\rightarrow\infty}\langle A\rangle_t~=~ \frac{\int d{\bf q}\int d\dot{{\bf q}}~
A({\bf q})\delta[H({\bf q},\dot{{\bf q}})-E]}
{\int d{\bf q}\int d\dot{{\bf q}}~\delta[H({\bf q},\dot{{\bf q}})-E]}.
\label{eq:muaverage}
\end{equation}
Per Eisenhart's metric, the average curvature $\kappa$ and the ratio $\rho\equiv\sigma/\kappa$, with $\sigma$ denoting the standard deviation of the curvature, are
\begin{equation}
\kappa~=~\frac{\left\langle\nabla^2V\right\rangle}{3N-1},\;\;\;
\rho={1\over\kappa}\frac{\sqrt{\left\langle(\nabla^2V)^2-
\langle\nabla^2V\rangle^2\right\rangle}}{\sqrt{3N-1}},
\label{eq:timecoef}
\end{equation}
in which $\nabla^2$ denotes the Laplacian $\partial_i\partial^i$, and $\rho$ corresponds physically to the ratio of the average curvature radius to the length scale of fluctuations~\cite{casetti95}.
By taking the curvature to vary randomly along a chaotic orbit, one can reduce Eq.~(\ref{eq:geodev}) to a stochastic-oscillator equation that can be solved analytically.
The solution yields an estimate of the largest Lyapunov exponent $\chi$:
\begin{eqnarray}
\chi(\rho)~&=&~\frac{1}{\sqrt{3}}
\frac{L^2(\rho)-1}{L(\rho)}\sqrt{\kappa};\nonumber\\
L(\rho)~&=&~\left[T(\rho)+\sqrt{1+T^2(\rho)}\right]^{1/3},\;\;
T(\rho)~=~\frac{3\pi\sqrt{3}}{8}\frac{\rho^2}{2\sqrt{1+\rho}+\pi\rho}.
\label{eq:analyticLyapunov}
\end{eqnarray}

The geometric quantities derive from the $6N$-dimensional microcanonical distribution.
Anticipating that granularity takes a long time to affect mixing, and wishing to identify conditions for rapid mixing, we now consider the influence of the 3-dimensional coarse-grained space-charge potential $V_s$ on a generic chaotic orbit.
The largest Lyapunov exponent for the coarse-grained system equates to the chaotic-mixing rate.
We presume the assumptions and approximations stated at the outset carry over to the coarse-grained system; the main justification is that the aforementioned previous work concerning low-dimensional autonomous Hamiltonians has shown the mixing rate in such systems usually depends only weakly on the dynamical details~\cite{kandrup01}.
We take the external focusing potential $V_f$ to be quadratic in the coordinates ${\bf x}$ comoving with the bunch, i.e., $V_f({\bf x})=({\bf\omega}\cdot{\bf x})^2/2$, wherein ${\bf \omega}=(\omega_x,\omega_y,\omega_z)$ corresponds to the focusing strength; the total potential is $V=V_f+V_s$.
Per Eq.~(\ref{eq:timecoef}) and Poisson's equation the quantities $\kappa$ and $\sigma$ are determined from $\nabla^2V=\omega_f^2-\omega_p^2({\bf x})$, in which $\omega_f^2=\omega_x^2+\omega_y^2+\omega_z^2$, $\omega_p^2({\bf x})=n({\bf x})q^2/(\epsilon_o m)$, $n({\bf x})$ is the (smoothed) particle density, $q$ and $m$ are the single-particle charge and rest mass, respectively, and $\epsilon_o$ is the permittivity of free space.
We then have
\begin{equation}
\kappa={1\over 2}\left[\omega_f^2-\langle\omega_p^2({\bf x})\rangle\right],\;\;\;
\rho={1\over\kappa}\frac{\sqrt{\langle n^2({\bf x})-
\langle n({\bf x})\rangle^2\rangle}}{n(0)\sqrt{2}}\;.
\label{eq:kappasigma}
\end{equation}
Inserting these results into Eq.~(\ref{eq:analyticLyapunov}) gives the associated time scale for chaotic mixing, $t_m\equiv 1/\chi$.
When the standard deviation of the density distribution is large, as can be the case when substructure is present, $\rho$ will be appreciable, and in turn Eq.~(\ref{eq:analyticLyapunov}) makes clear that $t_m$ will be a few space-charge-depressed periods $2\pi/\sqrt{\kappa}$.
Accordingly, the space-charge-depressed period, a quantity commensurate to the orbital period of a typical particle, constitutes a ``dynamical time'' $t_D$ for charged-particle beams.

To underscore the potential impact of collision{\it less} relaxation via chaotic mixing, it is of interest to compare $t_m$ to the collision{\it al} relaxation time $t_R$.
Perhaps the simplest way to develop an order-of-magnitude estimate of $t_R$ in a charged-particle bunch (a {\it non}neutral plasma) is to calculate the time required for a typical particle velocity to change by of order itself presuming collisions comprise a sum of incoherent binary interactions~\cite{rosenbluth}.
The result is $t_R/t_D\sim 0.1N/\mbox{ln}N$, wherein the Coulomb logarithm is conservatively taken to be $\mbox{ln}N$.
If we substitute plausible parameter values for real high-brightness beams, we find $t_R\gg t_D$; for example, $N=6.25\times 10^9$ (1 nC) gives $t_R\sim 10^7\;t_D$; hence, $t_m\ll t_R$ when chaotic mixing is prominent.
The remaining question is whether there is a significant population of globally chaotic orbits to mix, a question to which we now turn our attention.

\section{The Equations of Thermal Equilibrium\label{sec:TE}}
Consider a system, i.e., a bunch, of $N$ identical charged particles, e.g., electrons or protons.
For simplicity, invoke a Cartesian coordinate system whose origin lies at the bunch centroid.
Assume all particle velocities in this coordinate system are nonrelativistic.
The particles mutually interact via the Coulomb force and are confined by a static, externally applied, linear focusing force.
The focusing force may have different strengths along the three Cartesian axes.
Assume, apart from this focusing force, that the system is isolated and is in thermal equilibrium.
Accordingly, the total energy $E$ of each particle is conserved:
\begin{equation}
E~=~{1\over 2}mv^2+{1\over 2}m({\bf\omega}\cdot{\bf x})^2+q\phi({\bf x})\,;
\label{eq:energy}
\end{equation}
wherein $\bf{\omega}=(\omega_x,\omega_y,\omega_z)$ corresponds to the focusing strength; ${\bf x}=(x,y,z)$ denotes coordinates; $m$, $v$, and $q$ are the particle's rest mass, speed, and charge, respectively; and $\phi({\bf x})=(m/q)V_s$ is the space-charge potential arising from the collective Coulomb force.

To proceed, one would in principle work with the $6N$-dimensional microcanonical distribution of particles.
This distribution includes interactions at all scales, ranging from particle-on-particle to a single particle interacting with the bulk, smooth potential from all other particles.
Discreteness effects from $1/r^2$ particle collisions generate chaos~\cite{struckmeier96}; they cause nearby particle trajectories to separate exponentially.
The rate of exponential separation, i.e., the Lyapunov exponent, is an increasing function of $N$~\cite{hemsendorf02}.
In this sense, larger $N$ gives rise to more chaos.
However, the scale at which the separation saturates is a decreasing function of $N$.
Accordingly, in large-$N$, high-charge-density systems such as beams with space charge, discreteness establishes ${\it micro}$chaos~\cite{kansid01,sideris02,kansid02,Nbody}.
At the other extreme, that of a single particle interacting with the bulk, smooth potential, exponential separation of nearby chaotic particles (if any are present) saturates at a global scale, corresponding to a state of ${\it macro}$chaos.
Thus, initially nearby chaotic orbits evolve in three stages~\cite{kandrup02}: (1) very rapid exponential divergence that saturates at a scale large compared to the initial interparticle spacing but small compared to the system size; followed by (2) rapid exponential divergence that persists until the particles are globally dispersed; followed by (3) less rapid power-law divergence on a time scale $\propto(\ln N)t_D$, in which $t_D$ is a dynamical time commensurate to the orbital period.
If, in the smooth potential, the initially nearby particles execute regular motion rather than chaotic, then stage (2) is absent, and stage (3) proceeds on the much longer time scale $\propto(N^{1/2})t_D$~\cite{kansid02}.

Our interest here is in stage (2).
Specifically, we are concerned about the existence of, and time scale for, macroscopic chaos, i.e., chaotic mixing into the global region of phase space that is energetically accessible to the individual particles.
Accordingly, we specialize to the smooth 6-dimensional distribution function of a single particle, recognizing that discreteness effects vanish on macroscopic scales as the number density grows.
For the TE beam, this is just the Maxwell-Boltzmann distribution, $f({\bf x},{\bf v})\propto\exp(-H/kT)$, in which $H=E$ is the Hamiltonian, $k$ is Boltzmann's constant, and $T$ is the beam temperature.
The number density follows upon integrating over velocity space, and the space-charge potential follows upon solving Poisson's equation:
\begin{equation}
n({\bf x})=n(0)\exp\left[{-{1\over 2}m({\bf\omega}\cdot{\bf x})^2-q\phi({\bf x})}\over{kT}\right];\\
\label{eq:density}
\end{equation}
\begin{equation}
\nabla^2\phi({\bf x})=-{q\over\epsilon_o}n({\bf x}),~~~\phi({\bf x}=0)={\bf\nabla}\phi({\bf x}=0)=0,
\label{eq:potential}
\end{equation}
wherein $\epsilon_o$ is the permittivity of free space.

A much more convenient formulation arises by using dimensionless variables.
We introduce the Debye length $\lambda_{D0}$ and angular plasma frequency $\omega_{p0}$, both defined in terms of the centroid number density $n(0)$:
\begin{equation}
\lambda_{D0}^2\equiv{{\epsilon_o kT}\over{n(0)q^2}};~~~\omega_{p0}^2\equiv{{n(0)q^2}\over{\epsilon_o m}}.
\label{eq:norm}
\end{equation}
We then measure all lengths in the unit of $\lambda_{D0}$, i.e., ${\bf x}\leftrightarrow{\bf x}/\lambda_{D0}$, and all times in the unit of $1/\omega_{p0}$, i.e., $t\leftrightarrow\omega_{p0} t$.
In addition, we introduce the dimensionless potential $\Phi({\bf x})\equiv q\phi({\bf x})/(kT)$, and we normalize $n({\bf x})$ to the centroid density $n(0)$, i.e., $n({\bf x})\leftrightarrow n({\bf x})/n(0)$.
The number density and Poisson's equation then reduce to their dimensionless forms:
\begin{equation}
n({\bf x})=\exp\left[-{1\over 2}\Omega^2 R^2({\bf x})-\Phi({\bf x})\right];\\
\label{eq:normdensity}
\end{equation}
\begin{equation}
\nabla^2\Phi({\bf x})=-n({\bf x}),~~~\Phi({\bf x}=0)={\bf\nabla}\Phi({\bf x}=0)=0;
\label{eq:normpotential}
\end{equation}
wherein $\Omega^2\equiv(\omega_y/\omega_{p0})^2\leftrightarrow\omega_y^2$, and $R^2({\bf x})=(x/a)^2+y^2+(z/c)^2$, with the ``scale lengths'' $a$ and $c$ defined as $a\equiv\omega_y/\omega_x$ and $c\equiv\omega_y/\omega_z$.

Equations~(\ref{eq:normdensity}) and (\ref{eq:normpotential}) self-consistently provide the structure of the entire family of smoothed TE configurations.
The parameter $\Omega$ governs the strength of external focusing vis-$\grave{\mbox{a}}$-vis space charge.
The scale lengths $a$ and $c$ set the overall geometry: $a=c=1$ corresponds to spherical symmetry, $a=c\neq 1$ corresponds to cylindrical symmetry, and $a\neq 1$, $c\neq 1$, $a\neq c$ establishes a triaxial configuration.
The extreme case of maximum space charge corresponds to a density that is strictly uniform over the volume of the configuration, in which case $\Phi=-\Omega^2 R^2/2$ inside the beam.
Upon substituting into Poisson's equation, we see that the associated configuration carries the parameter $\Omega=\Omega_u=1/\sqrt{(1/a^2)+1+(1/c^2)}$.
This is the minimum permissible focusing strength; the bunch is unconfined if $\Omega<\Omega_u$, and the corresponding constraint on the parameter space is
\begin{equation}
{1\over a^2}+{1\over c^2}\geq{{1-\Omega^2}\over\Omega^2}\,.
\label{eq:constraint}
\end{equation}
Hence, the parameter set $[a,c;\Omega]$ fully specifies a TE configuration.

Upon solving for the space-charge potential $\Phi({\bf x})$, one can calculate orbits of test particles in the total potential.
Their trajectories follow from the (dimensionless) equation of motion:
\begin{equation}
{{d^2\bf x}\over{dt^2}}=-{\bf\nabla}\left[{1\over 2}\Omega^2R^2({\bf x})+\Phi({\bf x})\right].
\label{eq:motion}
\end{equation}
One can, of course, introduce arbitrary initial conditions for the orbits.
In our experiments, the initial condition on the velocity is ${\bf v}(0)=0$, and the total energy $E$ of a particle thereby corresponds to the potential energy associated with the initial position ${\bf x}(0)$.

A key challenge in exploring orbital dynamics throughout the parameter space is to integrate large numbers of orbits rapidly for sufficiently long evolutionary times.
Ideally, one would have analytic solutions for the density-potential pairs, from which the force on a particle at each time step can be quickly evaluated.
Unfortunately, the equations of equilibrium generally do not submit to analytic techniques.
Thus, in principle, one must solve these equations numerically, e.g., over a grid.
However, as delineated in the following section, it is possible to formulate approximate, semianalytic solutions, and these solutions enable a search of a broad range of the parameter space for regions that support chaotic orbits.
We now turn to that exploration.
Subsequently, for select cases, we compare these results against those derived from fully self-consistent numerical solutions.

\section{Approximate Solutions to the Equations of Equilibrium\label{sec:appx}}
A method to solve the equations of equilibrium is through a sequence of successive approximations~\cite{bohn83}.
A way to begin such a sequence is as follows:
(1) As a first approximation, represent the system as a configuration stratified on similar and similarly situated concentric ellipsoids.
A ``homeoid'' is a shell that is bounded by two similar and similarly situated concentric ellipsoids, and in which the surfaces of constant density are ellipsoids that are similar to, concentric with, and similarly situated with respect to the bounding ellipsoid.
Thus, the charge density is ``homeoidally striated'' in the first approximation (as is later illustrated in Fig.~\ref{fig:densitycompare}).
Determine the stratification by solving a spherically symmetric model of the equations of equilibrium.
(2) In the second approximation, derive the space-charge field corresponding to the homeoidally striated charge density, and then solve exactly the equations of equilibrium in this field.
(3 and up) Repeat the process until the density and potential converge.
In practice, one can carry out steps (1) and (2) of this recipe using semianalytic methods; to go further requires numerical techniques.

\subsection{Determination of the structure in the first approximation}
To invoke a spherically symmetric model of Eq.~(\ref{eq:normpotential}), we take the potential to be stratified over ellipsoids on which $R({\bf x})$ takes a constant value.
Then the spherically symmetric model corresponds to solving
\begin{eqnarray}
\nonumber
{1\over R^2}{d\over dR}\left(R^2{d\Phi_0(R)\over dR}\right)&=&-\exp\left[-{1\over 2}\Omega^2 R^2-\Phi_0(R)\right];\\
\Phi_0(0)&=&\left.{d\Phi_0\over dR}\right|_{R=0}=0.
\label{eq:poisson1}
\end{eqnarray}
This model defines the ``zeroth approximation'' $\Phi_0(R)$ to the potential.
In general Eq.~(\ref{eq:poisson1}) must be solved numerically; however, the solution is rapidly and easily accomplished with the aid of, e.g., a Runge-Kutta algorithm.

Once $\Phi_0(R)$ is determined, the corresponding homeoidally striated density becomes the first approximation to the number density:
\begin{equation}
n_1(R)=\exp\left[-{1\over 2}\Omega^2 R^2-\Phi_0(R)\right].
\label{eq:density1}
\end{equation}
By inspection~\cite{efe}, one can write down the space-charge potential corresponding to the number density $n_1(R)$, and this becomes the first approximation to the potential:
\begin{eqnarray}
\nonumber
\Phi_1({\bf x})&=&-{{ac}\over 2}\int_0^\infty{{du}\over\Delta(u)}\int_0^{R({\bf x};u)}dr\,r\,n_1(r)\\
&=&{{ac}\over 2}\int_0^\infty{{du}\over\Delta(u)}\left[r{{d\Phi_0(r)}\over{dr}}+\Phi_0(r)
\right]_{r=R({\bf x};u)}\,,
\label{eq:potential1}
\end{eqnarray}
wherein the second equality follows from an integration by parts, and the quantities $\Delta(u)$ and $R({\bf x};u)$ are
\begin{eqnarray}
\nonumber
\Delta(u)&=&\sqrt{(a^2+u)(1+u)(c^2+u)}~,\\
\nonumber
R({\bf x};u)&=&\sqrt{{x^2\over{a^2+u}}+{y^2\over{1+u}}+{z^2\over{c^2+u}}}~.
\label{eq:auxiliary}
\end{eqnarray}
Hence, in the first approximation the number density is homeoidally striated, but the space-charge potential is not.

\subsection{Determination of the structure in the second and higher approximations}
The number density in the second approximation, $n_2({\bf x})$, follows upon substituting $\Phi_1({\bf x})$ calculated from Eq.~(\ref{eq:potential1}) into Eq.~(\ref{eq:normdensity}).
For the special case of spherical symmetry, all orders of approximation agree with one another, but this is of course not true for a general triaxial geometry.
To go further requires numerical methods, e.g., solving Poisson's equation for the potential $\Phi_2$ corresponding to the density $n_2$, substituting the result into Eq.~(\ref{eq:normdensity}) to obtain $n_3$, and successively repeating the process until convergence is achieved.
As discussed in Sec.~\ref{sec:exact} below, we use a different method, a multigrid algorithm, for solving Eqs.~(\ref{eq:normdensity}) and (\ref{eq:normpotential}) numerically.

\section{Survey of the parameter space\label{sec:survey}}
Gathering sufficient data to support precise, statistically based conclusions concerning orbital behavior in a given potential requires integrating thousands of orbits in that potential.
And before these orbits can be tracked, the potential needs to be ascertained to sufficient accuracy.
In principle, and for each choice of parameters, one must construct the ``exact'' potential $\Phi({\bf x})$ by numerically solving the corresponding Poisson equation.
This can be a computationally tedious process, and the solution is by necessity defined over a grid.
Next, orbit integration through the grid requires accurate interpolation to evaluate the potential and corresponding particle acceleration between grid points.
For sufficient resolution, the time steps need to be appropriately small; accordingly, many interpolations are required, and integrating many orbits is computationally time-consuming.
This process is feasible for studying a few choices of parameter sets, and it underlies the results of Sec.~\ref{sec:exact} below.
However, to survey the entire parameter space, i.e., to investigate many choices of parameter sets, the process becomes prohibitive.
For this purpose one must resort to using approximate potentials.

Sec.~\ref{sec:appx} above details a sequence of approximations, the first elements of which are semianalytic.
The zeroth-order potential $\Phi_0$, derived from Eq.~(\ref{eq:poisson1}), is easy to evaluate, and it enables fast, high-precision orbital integration.
However, $\Phi_0$ itself may be a crude approximation to the exact potential; the approximation gets progressively worse as the parameter sets deviate further from spherical symmetry.
One might expect the potential $\Phi_1$ of the first approximation to provide a better model.
However, its underlying integral, given in Eq.~(\ref{eq:potential1}), adds additional complexity and time to the orbit integrations.
We tried evaluating this integral at each time (thus position) step along the orbit, but doing so made the orbit computations prohibitively long.
The alternative is to evaluate the integral over a grid and then do orbit integrations through the grid.
As previously mentioned, integrations through a grid are too computationally expensive to enable a parameter survey.
Moreover, if one is able to solve Poisson's equation for the exact potential $\Phi({\bf x})$, then there is neither computational benefit nor motivation for using $\Phi_1$.
Our strategy is to explore a few choices of parameter sets in the exact potential to strengthen conclusions from our survey, and this necessitated developing the Poisson solver described in Sec.~\ref{sec:exact}.
For these reasons, we use the potential of the zeroth approximation, $\Phi_0$, to survey the parameter space.
For a few specific parameter sets for which the results of the zeroth approximation look especially interesting, we then check the results using the numerically evaluated exact potential, $\Phi({\bf x})$.

\subsection{Solution for the zeroth-order potential\label{subsec:zeroth}}

Per Eq.~(\ref{eq:constraint}), the minimum possible focusing strength corresponding to a spherically symmetric system is $\Omega=\Omega_u=1/\sqrt{3}$.
Accordingly, we choose focusing strengths in keeping with the following labeling convention:
\begin{equation}
\Omega_i={1\over{\sqrt{3}}}\left[1+10^{1-i}-10^{1-I_{max}}\right]\,,
\label{eq:case}
\end{equation}
with $i=1,2,...,I_{max}$ and $I_{max}=10$.
Choices of parameters $[a,c;\Omega_i]$ then must be selected based on the constraint of Eq.~(\ref{eq:constraint}) which bounds the parameter space of exact potentials.
Note, however, that we can examine any desired geometry: oblate axisymmetric (for which $a=1,~c<1$), prolate axisymmetric (for which $a=c<1$), and the full range of oblate-through-prolate triaxial systems.
Hereafter we refer to ``Case 1, Case 2,...'' according to ``$i=1,2,...$'' in Eq.~(\ref{eq:case}), respectively.

\begin{figure}
\includegraphics[width=15cm]{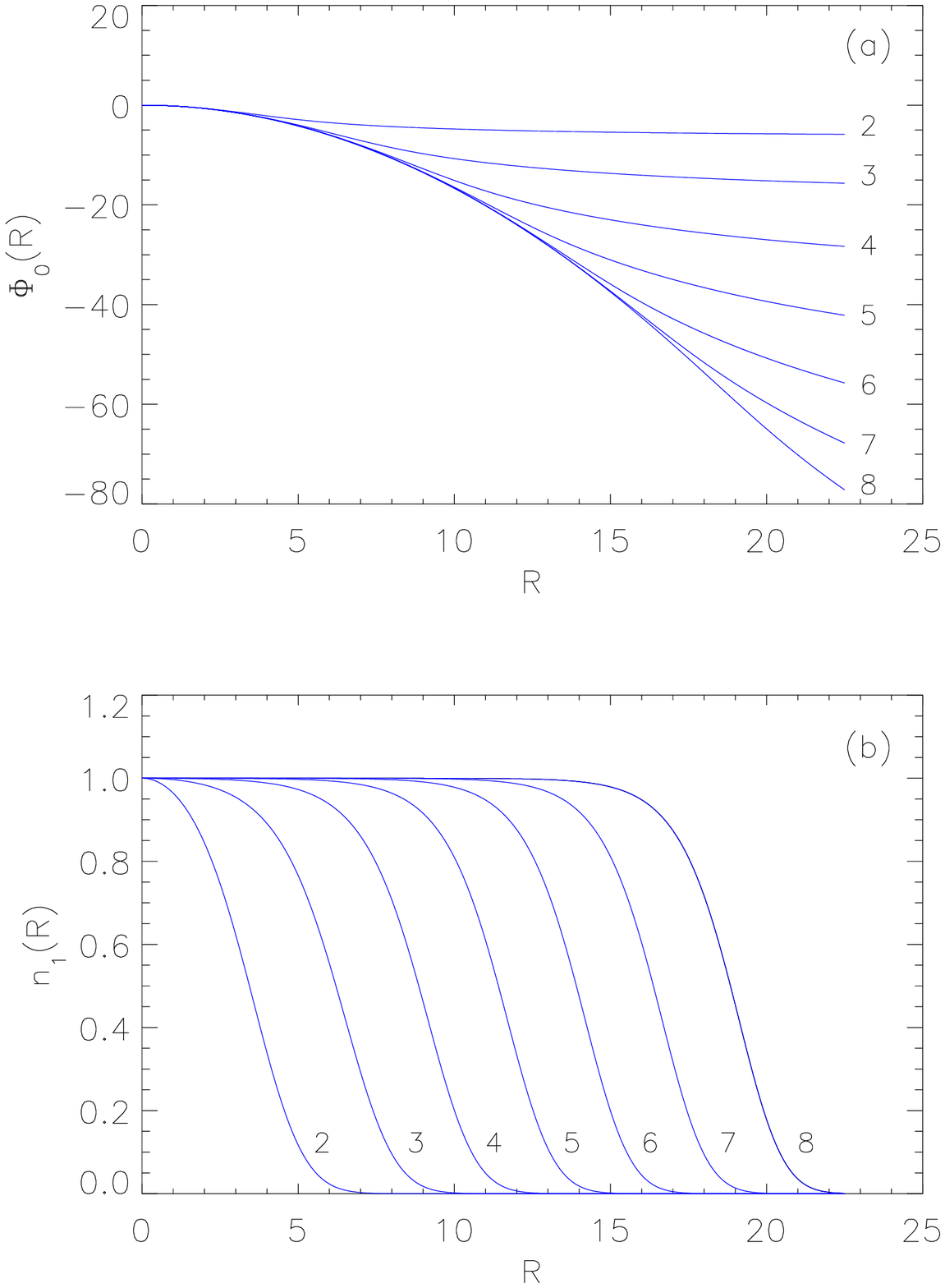}
\caption{\label{fig:zeroth}(a) Space-charge potential $\Phi_0~\mbox{vs.}~R$ for $\Omega$ corresponding to Cases 2-8 as defined in Eq.~(\ref{eq:case}).
(b) Number density $n_1~\mbox{vs.}~R$ for Cases 2-8.}
\end{figure}

Plots of $\Phi_0(R)$ versus $R$ derived from Eq.~(\ref{eq:poisson1}) appear in Fig.~\ref{fig:zeroth}.
Also shown are the corresponding profiles of the number densities $n_1(R)$ constructed in the first approximation.
For larger ``case numbers'' $i$, the density contains larger quasi-uniform central regions.
In the outer regions the density decreases, over a length commensurate to the Debye length, to a low-density tail.
The space-charge force in the quasi-uniform ``core'' is correspondingly quasi-linear; however, it is manifestly nonlinear in the ``Debye fall-off region'' (henceforth called the ``Debye tail'').
Fig.~\ref{fig:zeroth} shows that the choices of $\Omega_i$ per Eq.~(\ref{eq:case}) span a wide range of space charge.
At the one extreme, zero space charge, the density profile is gaussian.
Then, the range of $i$ spans from small space charge ($i=1$) for which the density profile is approximately gaussian, through the fully space-charge-dominated, uniform beam ($i=I_{max}=10$) for which $\Omega=1/\sqrt{3}$.
Note that Case 5 ($\Omega\simeq 1.0001/\sqrt{3}$) represents ``intermediate space charge''; the density falls off over a length scale comparable to that of the core.

\subsection{Methodology for orbital analysis\label{subsec:method}}
\subsubsection{Samples of orbits, power spectra, and complexity\label{subsubsec:samples}}
After choosing a parameter set $[a,c;\Omega_i]$ and solving for $\Phi_0[R({\bf x})]$, we began by generating 2000 initial coordinates uniformly spanning the volume occupied by the core and Debye tail of the density $n_1[R({\bf x})]$.
The initial velocities were all chosen to be zero.
Next, starting with these initial conditions, for every orbit we integrated the equations of motion, cf. Eq.~(\ref{eq:motion}), using $\Phi=\Phi_0$ for at least 100 orbital periods and, in most cases, for $\agt 200$ orbital periods.
The integrations were done using a fifth-order Runge-Kutta algorithm~\cite{ptvf} with variable time step.
As the integration proceeded, we computed the largest short-time Lyapunov exponent of each orbit using a well-established algorithm in the field of chaotic dynamics~\cite{bennetin}.
The idea is to evolve two initial conditions that start from a very close distance for about one dynamical time, then renormalize to bring the two particles close together again, and repeat the process until the average exponent associated with the orbital separation converges to an almost stable value.
Typically convergence was achieved within $\sim 100$ orbital periods.

After computing the orbits, we extracted the power spectrum for each orbit using a fast-Fourier-transform algorithm~\cite{ptvf}.
In doing so, we recorded each orbit at a rate $\sim 40$ times per orbital period.
From the spectrum we computed the total power.
Then we sorted the spectral frequencies in descending order, and starting from the highest frequency we added as many frequencies as were needed to reach 90\% of the total power.
The required number of frequencies is defined to be the ``complexity'' $n$ of the orbit~\cite{complexkandrup}.

\subsubsection{Criterion for chaos\label{subsubsec:criterion}}
\begin{figure}
\includegraphics[width=15cm]{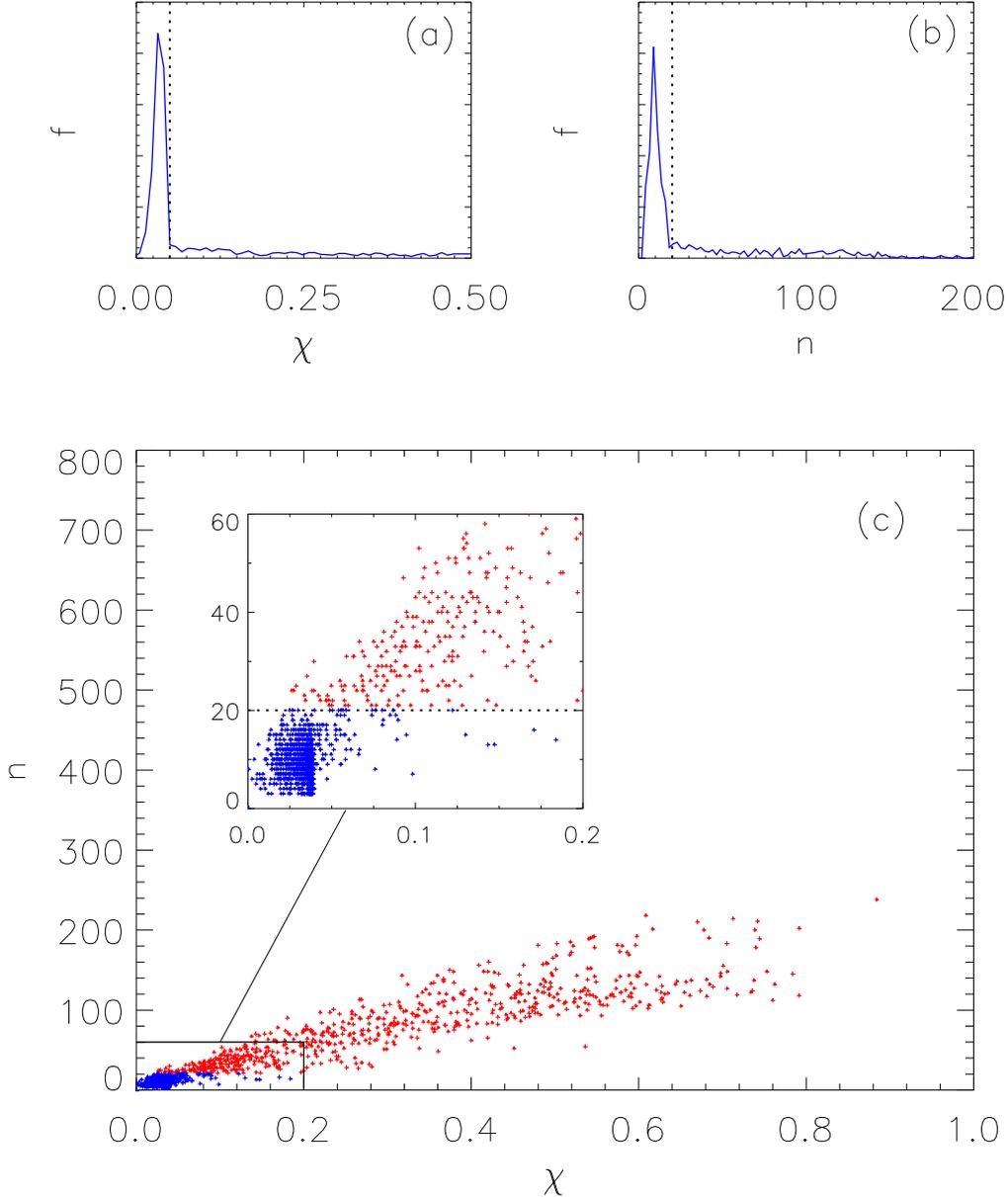}
\caption{\label{fig:measure}(a) Distribution $f(\chi)$, in arbitrary units, of Lyapunov exponents $\chi$ (Case 5, $a^2=0.5,~c^2=1.5$) in the zeroth approximation $\Phi_0$.
The unit of $\chi$ is $t_D^{-1}$.
The distribution peaks at low values of $\chi$ for which the respective orbits are regular.
The dotted line, hand-drawn at the point where the distribution levels off, suggests one possible criterion of chaos.
(b) Distribution $f(n)$ of complexities $n$ corresponding to the orbits of the first panel, which looks qualitatively similar to $f(\chi)$.
(c) Complexities $n$ versus Lyapunov exponents $\chi$.
The inset reveals that the concentration of regular orbits near the origin lies inside sharply defined boundaries.}
\end{figure}

Our first and foremost interest is to determine how many of the 2000 orbits in our sample are chaotic in a given TE configuration.
Accordingly an objective, quantitative criterion for chaos is needed.
There is no universally accepted criterion; hence, we developed our own using the following rationale.
Both the largest short-time Lyapunov exponent $\chi$ and the complexity $n$ are well-established, conventional measures of chaos~\cite{tabor}.
A first piece of information for defining the criterion comes from plotting $n$ versus $\chi$ for all of the orbits.
Fig.~\ref{fig:measure} provides an example. 
It shows that (a) $n$ increases approximately linearly with $\chi$, as one might expect since both quantities are measures of chaos, and (b) the regular orbits occupy a sharply defined region close to the origin of the $n$-vs.-$\chi$ plot.
The borders of this ``region of regularity'' thereby offer three possible criteria for chaos: one involving only the Lyapunov exponent, one involving only the complexity, and one involving both quantities.

We chose to base our criterion only on the complexity for the following reason.
Though no problem arises in computing Lyapunov exponents in the zeroth approximation, such is not the case in the exact potential, wherein we found that longer integration times are required to achieve adequate convergence.
Recall that the exact potential must be specified over a three-dimensional grid.
Accordingly, interpolation errors and discontinuities between the cells can affect computations of Lyapunov exponents because they involve the distance between two initially nearby orbits, which is a local property that is sensitive to the grid size and the order of interpolation.
By contrast, a computation of complexity, in that it involves the Fourier spectrum of an individual orbit, avoids reference to nearby orbits and is thus a global property influenced little by the grid size, a notion that we corroborated during the course of our numerical studies.
For simulations in exact potentials, we chose a grid size and interpolation algorithm (cf. Sec.~\ref{sec:exact} below) such that numerical errors had negligible effect on the computation of individual orbits.
A standard measure of the ``goodness'' of an orbital integration is the degree to which total energy is conserved~\cite{orbitref}; for every orbit we achieved conservation of total energy with relative error $\leq 10^{-6}$.
This is some two orders of magnitude better than contemporary standard practice.

Investigations of $n$-vs.-$\chi$ plots for many zeroth-order TE potentials led us to a specific quantitative criterion, namely, an orbit is categorized as chaotic if its complexity $n>20$ [cf. Fig.~\ref{fig:measure}(c)].
This choice was also checked and confirmed by carefully inspecting hundreds of plots of individual orbits in several geometries.

\begin{figure}
\includegraphics[width=13cm]{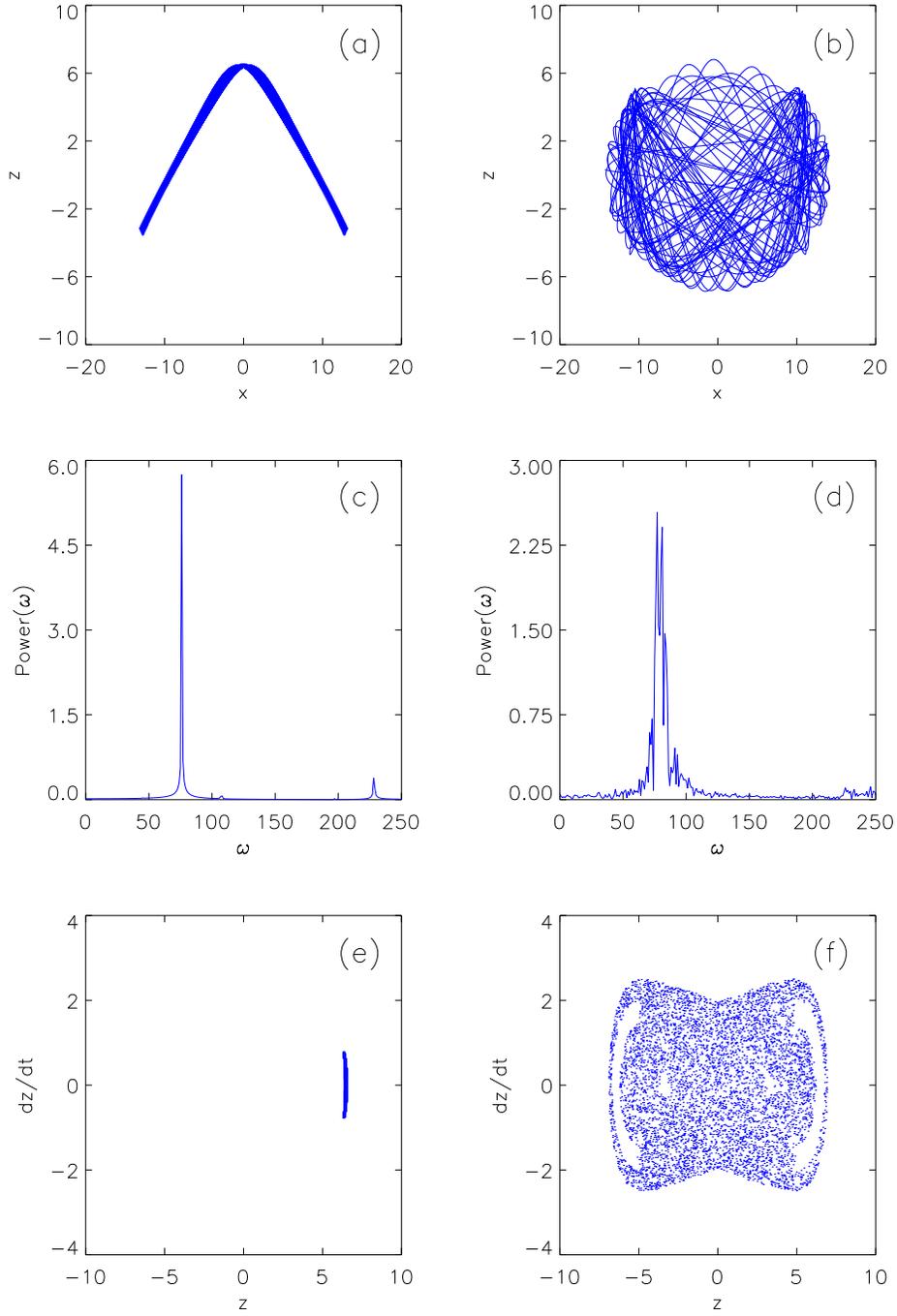}
\caption{\label{fig:orbits} Two representative orbits in the potential $\Phi_0$ of the zeroth approximation (Case 5, $a^2=1.0,~c^2=0.5$), one regular (a,c,e) for which $n=8$ and $\chi=0.028\;t_D^{-1}$, and one chaotic (b,d,f) for which $n=178$ and $\chi=0.640\;t_D^{-1}$, each with similar total energies: (a,b) orbits on (x,z)-plane; (c,d) power spectra in $x$-direction; (e,f) surfaces of section $dz/dt~\mbox{vs.}~z$.}
\end{figure}

Plots of two representative orbits, one regular and one chaotic, as well as their power spectra in the $x$-direction and their surfaces of section in the $dz/dt\mbox{-vs.-}z$ phase space, appear in Fig.~\ref{fig:orbits}.
Both orbits have similar total energies and evolve in $\Phi_0$ corresponding to Case 5 and $a^2=1.0,~c^2=0.25$, a strongly oblate spheroid.
The first orbit has $n=8$ and $\chi=0.028\;t_D^{-1}$; the second orbit has $n=178$ and $\chi=0.640\;t_D^{-1}$.
Their differences are striking.
One major difference concerns their power spectra.
The regular orbit exhibits two very distinct frequencies in its spectrum, whereas the chaotic orbit features a near-continuum spanning a large number of frequencies.
The surfaces of section are computed by recording $dz/dt$ and $z$ when $x=y=0$. The regular orbit is seen to stay within a localized region of the $dz/dt$-vs.-$z$ phase space, whereas the chaotic orbit largely fills a global area of phase space commensurate to its total energy.

\begin{figure}
\includegraphics[width=15cm]{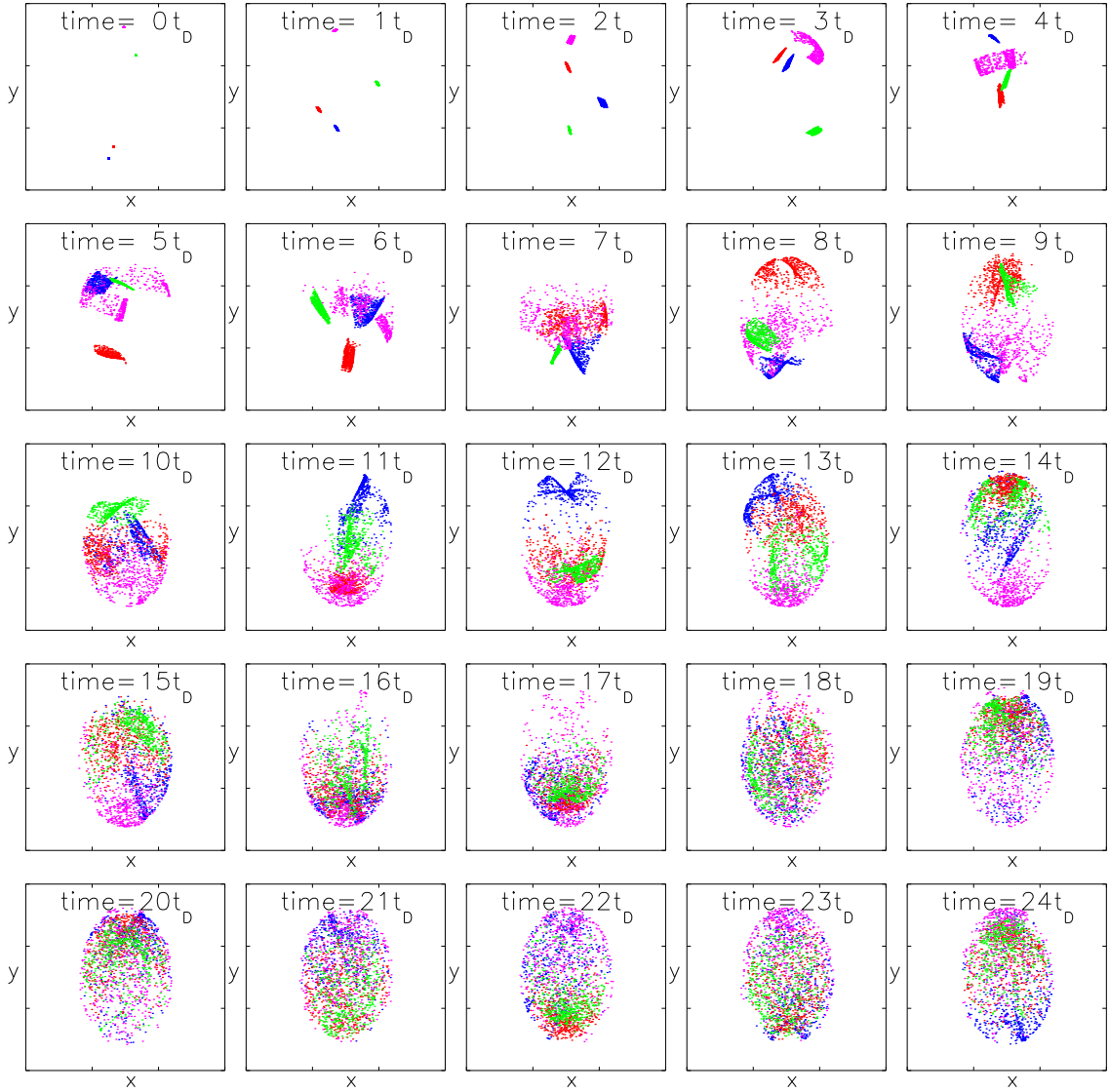}
\caption{\label{fig:clumps} Snapshots of four different clumps of chaotic orbits (Case 5, $a^2=0.5$, $c^2=1.5$) evolving in the potential $\Phi_0$.
Each clump initially occupies a cube of volume $0.05^3$, but exponentially grows to fill a volume commensurate to the total particle energy.}
\end{figure}

Results of a numerical experiment that highlights the largest Lyapunov exponent, i.e., the rate of global chaotic mixing, appears in Fig.~\ref{fig:clumps}.
Here, four initially localized clumps comprising 500 chaotic orbits evolve in the zeroth-order potential $\Phi_0$ corresponding to Case 5 and $a^2=0.5,~c^2=1.5$, a triaxial configuration.
Each clump initially occupies a cube of size $0.05^3$ in the configuration space.
The figure reveals that each clump mixes through a global region of phase space with an e-folding time comparable to a dynamical time $t_D$, taken here to be the orbital period corresponding to the total energy of the individual particles comprising the clump.
After some tens of $t_D$ each clump has spread through a volume commensurate to the total particle energy. 

\subsection{Survey results\label{subsec:surveyresults}}
Our strategy for surveying the parameter space of the TE configurations is as follows.
We conduct the survey using the zeroth-order potential $\Phi_0$ found from Eq.~(\ref{eq:potential1}).
Recall that this potential depends on the focusing strength $\Omega$ and, through $R({\bf x})$, the scale lengths ($a,~c$); we thus choose a specific parameter set $[a,c;\Omega]$.
Then we integrate 2000 initial conditions generated by uniformly sampling (1) the volume spanned by the configuration, and afterward (2) the volume spanned only by the Debye tail.
Having established the criterion for chaos (complexity $n>20$), we identify and count the chaotic orbits in the respective sample, and we express this number as a percentage of the sample.
This percentage may be viewed as an indication of the extent to which a given parameter set supports globally chaotic orbits.
However, it should not be taken too literally in that the initial conditions are distributed uniformly through a volume; they are not weighted by the actual density distribution.

Our numerical experiments fall largely into two categories.
Category I pertains to keeping $\Omega$ fixed to its Case 5 value (intermediate space charge), i.e., $i=5$ in Eq.~(\ref{eq:case}), and then varying $a$ and $c$ within the constraint of Eq.~(\ref{eq:constraint}).
Category II pertains to keeping $c^2=0.5$ fixed, and then varying $\Omega$ and $a$.

\begin{figure}
\includegraphics[width=15cm]{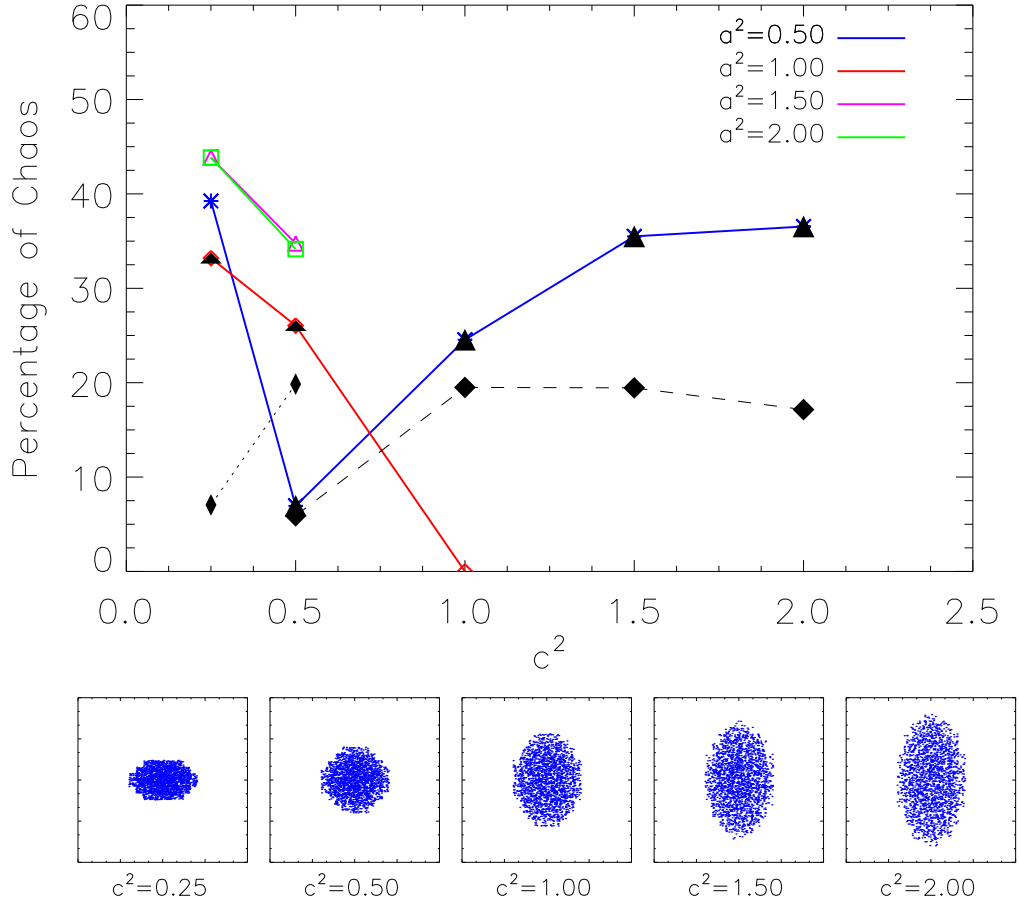}
\caption{\label{fig:catIa} Top: Percentage of chaotic orbits for Case 5 vs. $c^2$ in the potential $\Phi_0$ with different choices of $a^2$.
The initial conditions uniformly sample the region $0\le R\le 15$, which in essence covers the volume spanned by Case 5 configurations [cf. Fig.~\ref{fig:zeroth}(b)].
Results derived from exact solutions $\Phi({\bf x})$ of Poisson's equation are also plotted for comparison; they are joined by a dotted line ($a^2=1,~c^2<1$) or by dashed lines ($a^2=0.5$).
These exact results are to be compared to their zeroth-order counterparts delineated with blackened symbols to aid the eye.
Bottom: Initial conditions, projected onto the $(x,z)$-plane, used for $a^2=0.5$; $z$-axis is vertical.}
\end{figure}

\begin{figure}
\includegraphics[width=15cm]{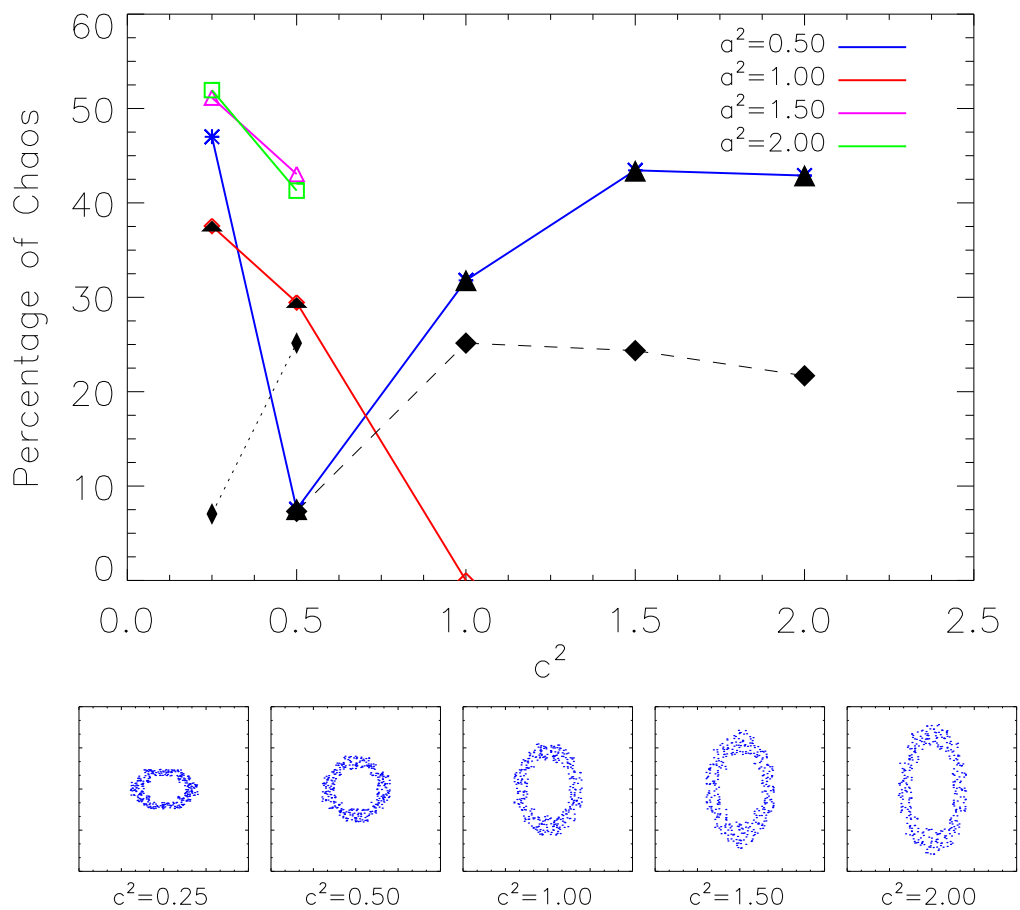}
\caption{\label{fig:catIb} Same as Fig.~\ref{fig:catIa}, but with initial conditions that uniformly sample only the Debye tail $9\le R\le 15$.
For ease of visualization, the bottom panel shows only the initial conditions for which $y\sim 0$.}
\end{figure}

Figures~\ref{fig:catIa} and~\ref{fig:catIb} depict the percentage of chaotic orbits for a portion of the Category I experiments.
Also shown in the bottom panels of these figures are the initial conditions, projected onto the $(x,z)$-plane, corresponding to the axisymmetric configurations for which $a^2=0.5$.
In addition, results from analogous experiments in exact potentials are plotted for comparison.
The figures exhibit a number of features.
First and foremost is the indication that most of the configurations support a considerable population of chaotic orbits.
This is true even for axisymmetric configurations, but of course it is not true for spherically symmetric configurations in that their potentials are integrable.
Second, orbits for which the initial conditions all lie within the Debye tail reflect a higher percentage of chaos than were they distributed through the entire configuration space.
Third, prolate axisymmetric configurations (for which only one datum is shown in each of these figures) support little chaos.
Fourth, for reasons to be elaborated in Sec.~\ref{sec:exact} below, the exact potentials support less chaos than their zeroth-order counterparts.

\begin{figure}
\includegraphics[width=16cm]{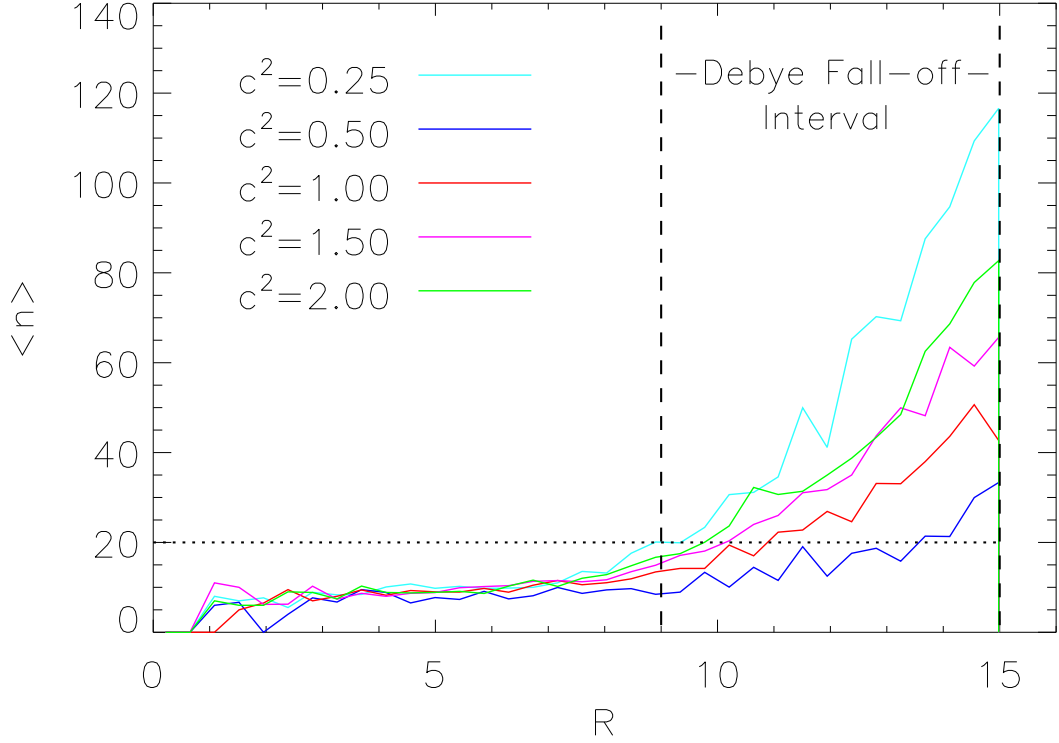}
\caption{\label{fig:catIdebye} Average complexity $\langle n\rangle$ versus homeoidal coordinate $R$ for Case 5 in the potential $\Phi_0$ with $a^2=0.5$ and different choices of $c^2$.
Essentially all of the chaotic orbits reach into the Debye tail.}
\end{figure}

Further analysis reveals that, for configurations in which they are present, essentially all of the chaotic orbits originate in the Debye tail.
Figure~\ref{fig:catIdebye} dramatically illustrates this finding.
To assemble this figure, the initial conditions are sorted and binned into increments of $R$ spanning 0.1 units of length.
Then, in each increment, the complexity $n$ of every orbit is computed, and the complexities are averaged to obtain $\langle n\rangle$.
The process is repeated for all of the increments, and then for many different parameter sets.
The results, $\langle n(R)\rangle$ versus $R$ for Case 5 with $a^2=0.5$ and several choices of $c^2$, comprise Fig.~\ref{fig:catIdebye}.

\begin{figure}
\includegraphics[width=15cm]{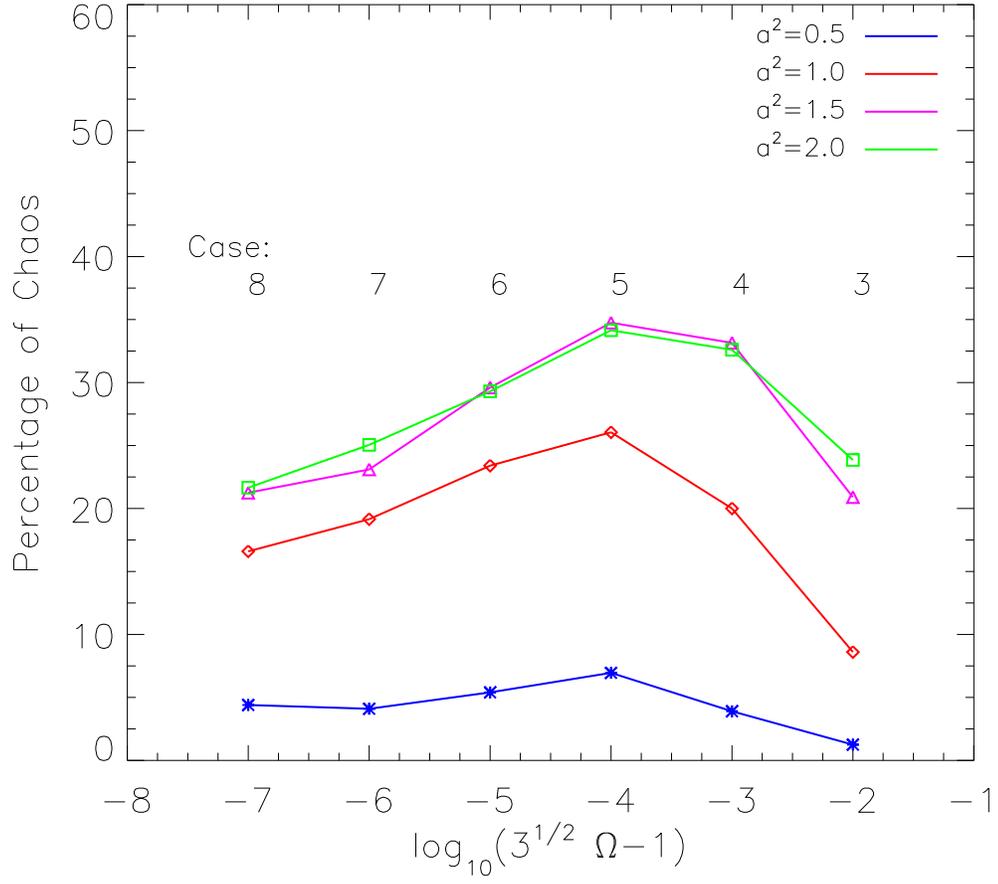}
\caption{\label{fig:catII} Percentage of chaotic orbits in the potential $\Phi_0$ with $c^2=0.5$ and $\Omega$ corresponding to Cases 3-8 as defined in Eq.~(\ref{eq:case}).
Different curves correspond to different choices of $a^2$.
The initial conditions uniformly sample the respective configurations.}
\end{figure}

Figure~\ref{fig:catII} depicts the percentage of orbits for the Category II experiments for which the initial conditions are uniformly distributed over the volume spanned by the configuration.
Cases corresponding to intermediate space charge would seem to support more chaos.
This finding makes sense when juxtaposed against the limiting cases of zero space charge at one extreme and zero Debye tail, i.e., the uniform beam, at the other extreme.
With zero space charge, only the linear forces of the external potential influence the particles.
For the uniform beam the external and space-charge potentials cancel one another so that the particles move freely, apart from reflections at the boundary of the configuration.
In both extremes the orbits are all regular, excepting billiard effects, if any, associated with shapes of boundary surfaces of uniform bunches.
The figure also suggests that prolate, axisymmetric TE configurations support little chaos.

\section{Numerical experiments in exact potentials\label{sec:exact}}
As a matter of principle, one must be concerned about the extent to which subtle structure in the potential can influence the qualitative behavior, and in particular the chaoticity, of an orbit.
One well-known example is that of the Toda potential; the full Toda potential is integrable and supports only regular orbits, but generally a truncated Toda potential is not integrable and supports a population of chaotic orbits~\cite{kandrup94}.
Our survey of the parameter space of TE configurations centered on the use of $\Phi_0$, a generally crude approximation to the true potential.
The survey suggests a large region of the parameter space supports sizeable populations of chaotic orbits, wherein all of these orbits reach into the Debye tail.
We may expect in general that the density profile, particularly that of the Debye tail, corresponding to the exact potential is considerably different from that corresponding to $\Phi_0$.
For example, in the limit of distances very far from the centroid, the exact space-charge potential will approach spherical symmetry, whereas $\Phi_0$ is everywhere homeoidally striated.
Accordingly, to check and have confidence in the qualitative results of Sec.~\ref{sec:survey}, we must repeat the numerical experiments in a suitably broad collection of exact potentials.
As mentioned earlier, the reason we did not base the survey on exact potentials is that the respective numerical experiments are computationally expensive. 

To integrate Eq.~(\ref{eq:normpotential}) governing the exact potential $\Phi({\bf x})$, which is a fully three-dimensional partial differential equation (PDE), we chose a multigrid algorithm~\cite{brandt}.
The algorithm requires boundary conditions be specified over the surface of the volume occupied by the grid.
We chose a cubic grid volume greatly exceeding the volume of interest, i.e., that spanning the Debye fall-off of the density.
Then we calculated the boundary conditions over the surface of this volume using the formalism of the first approximation, specifically, Eq.~(\ref{eq:potential1}).
Because the resulting boundary conditions are only first approximations to the true boundary conditions, we checked our numerical solutions by varying the positions of the bounding surfaces of the grid by factors of five, and we found negligible change in the results over the volume of interest.
Applying the multigrid algorithm to three dimensions involves nontrivial manipulations of the inherent restriction, interpolation, and relaxation routines.
In the process, a nonlinear algebraic equation emerges due to the nonlinearity of the PDE.
It was solved using an iterative method that combines Newton-Raphson and bisection techniques~\cite{ptvf}.

After tabulating the exact potential in three dimensions, we used a fifth-order Runge-Kutta algorithm with variable time step to evolve the individual orbits in time, within which the force at every time (thus position) step was computed using a three-dimensional interpolation scheme.
The resulting orbits conserved total energy with relative error better than $10^{-6}$ and sometimes as low as $10^{-7}$.

\begin{figure}
\includegraphics[width=15cm]{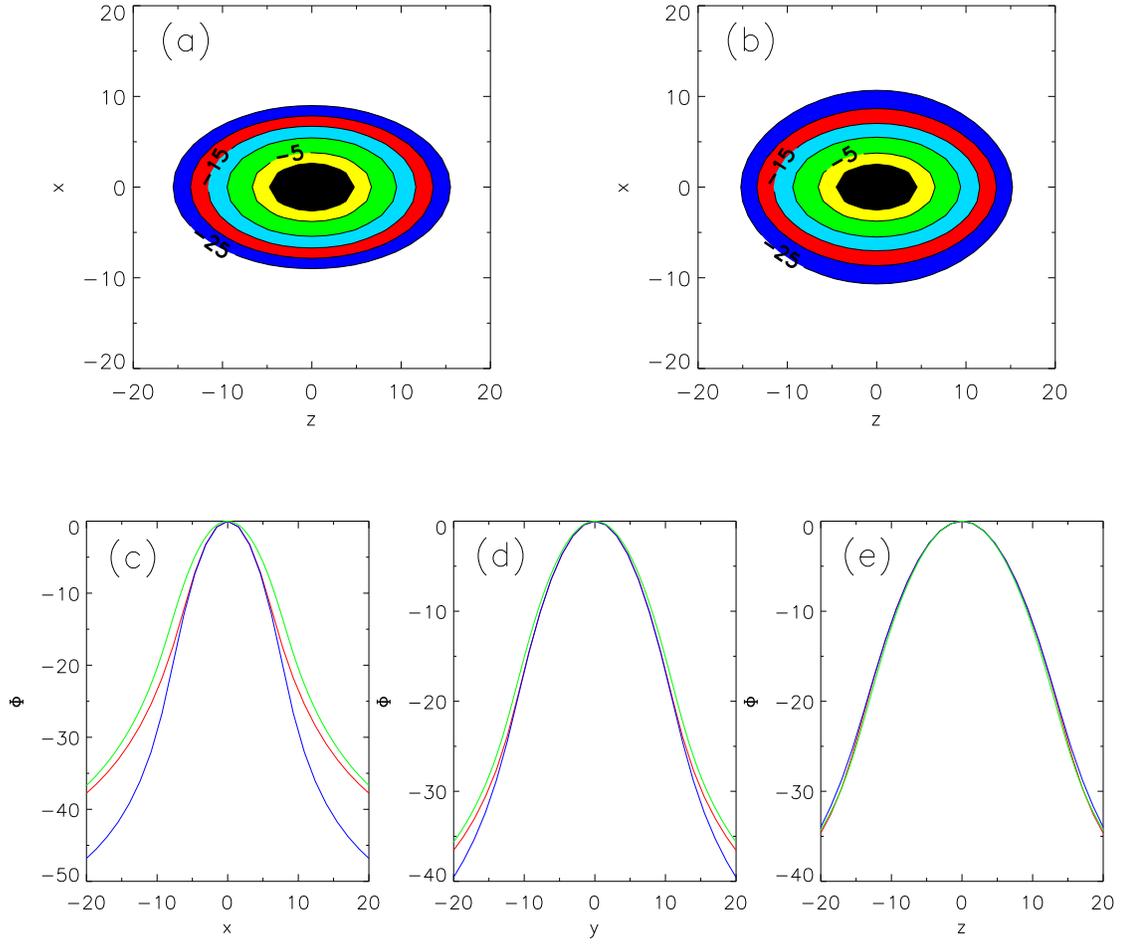}
\caption{\label{fig:potcompare1} Comparisons between the space-charge potential $\Phi_0$ in the zeroth approximation and the exact potential $\Phi({\bf x})$ (Case 5, $a^2=0.5,~c^2=1.5$): (a) isopotential contours of $\Phi_0$, (b) isopotential contours of $\Phi({\bf x})$, (c,d,e) profiles along the ($x,y,z$)-axes, respectively (blue and red curves pertain to $\Phi_0$ and $\Phi({\bf x})$, respectively).}
\end{figure}

\begin{figure}
\includegraphics[width=15cm]{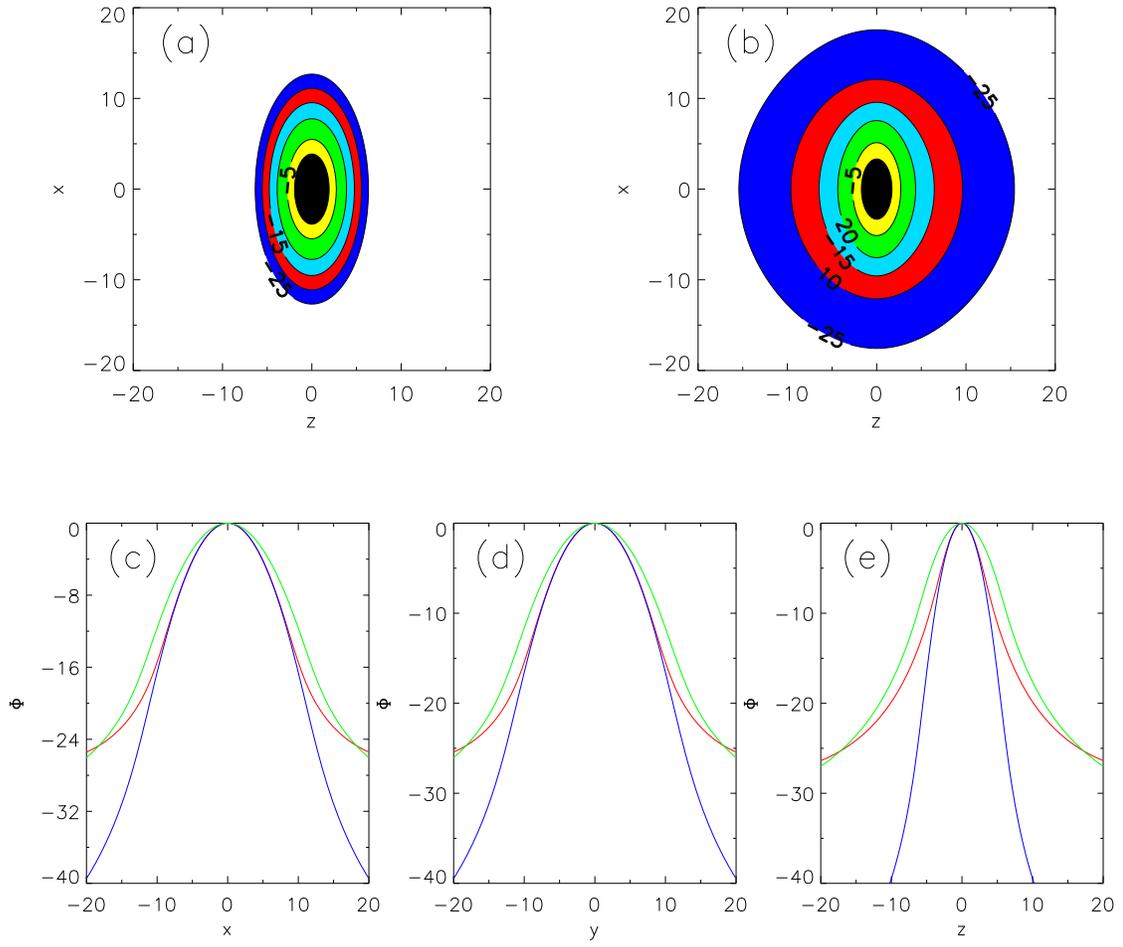}
\caption{\label{fig:potcompare2} Same as Fig.~\ref{fig:potcompare1}, but with $a^2=1.0,~c^2=0.25$.}
\end{figure}

Figure~\ref{fig:potcompare1} exemplifies the difference between the zeroth-order and exact potentials.
The top two panels present isopotential contours in the $(x,z)$-plane for Case 5 with $a^2=0.5,~c^2=1.5$, a triaxial configuration.
The bottom panels show how the two potentials compare along each of the ($x,y,z$)-axes.
Obviously there are, and there should be, differences, but the important question is to what extent these differences alter the qualitative evolution of the orbits and, in turn, the complexities that characterize them?
Analogous graphs for Case 5 and $a^2=1.0,~c^2=0.25$, a strongly oblate spheroid, appear in Fig.~\ref{fig:potcompare2}, from which the respective differences are seen to be much more pronounced.

\begin{figure}
\includegraphics[width=15cm]{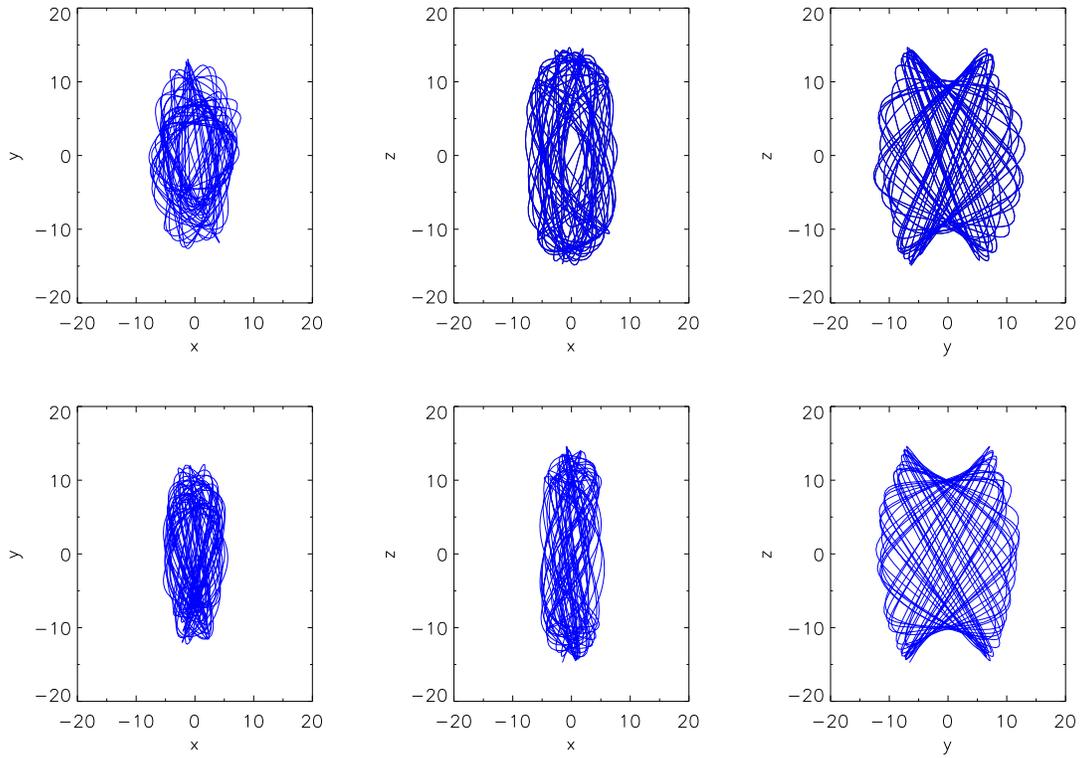}
\caption{\label{fig:orbitcompare} A chaotic orbit in the zeroth-order potential (top panels) and exact potential (bottom panels) of Fig.~\ref{fig:potcompare1} evolved from the same initial conditions.
The orbit is similar, but not identical, in the two potentials.}
\end{figure}

\begin{figure}
\includegraphics[width=15cm]{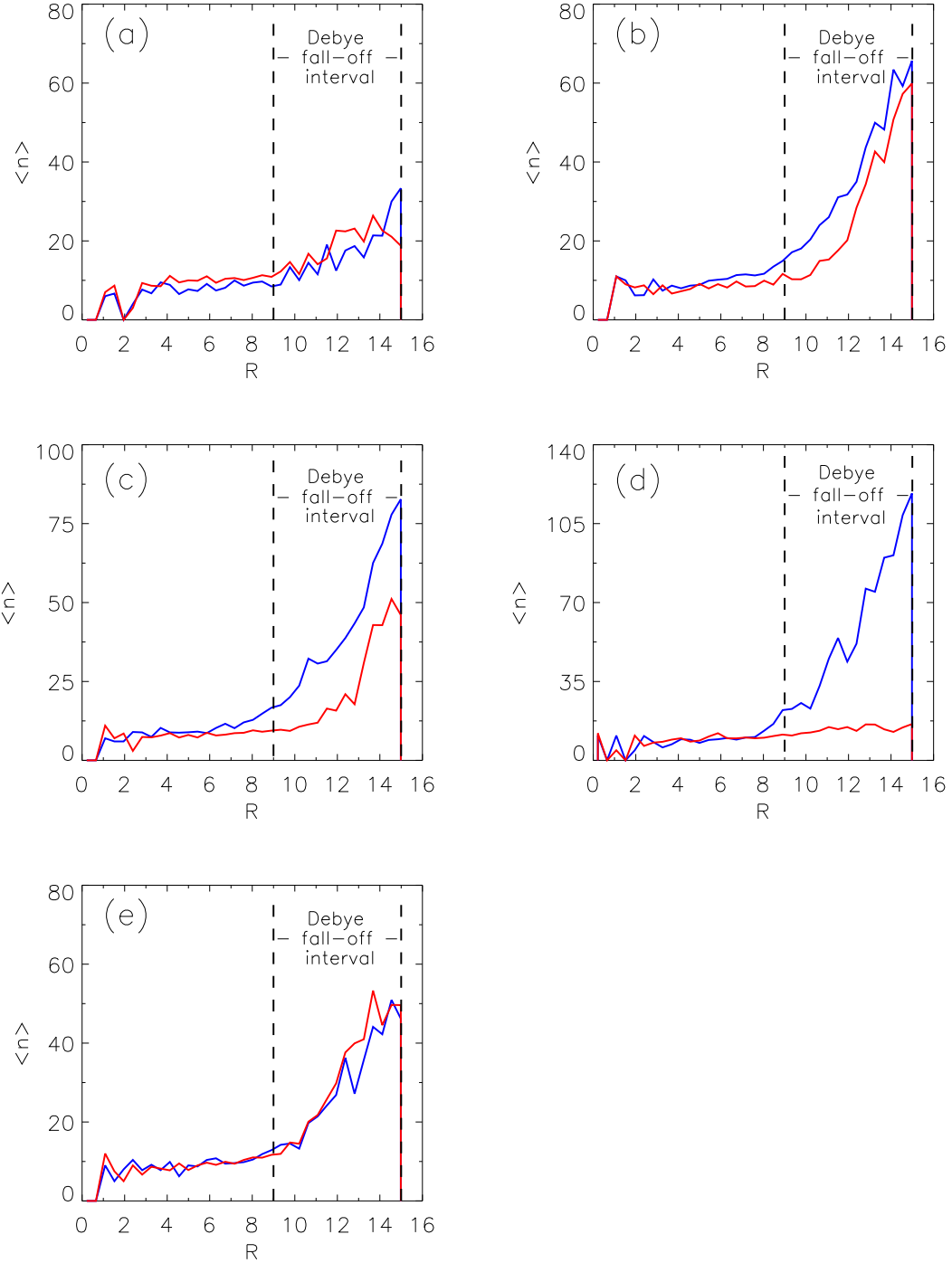}
\caption{\label{fig:complexitycompare} Average complexity $\langle n\rangle$ versus homeoidal coordinate $R$ for zeroth-order (blue) and exact (red) Case 5 potentials with: (a) $a^2=0.5,~c^2=0.5$; (b) $a^2=0.5,~c^2=1.5$; (c) $a^2=0.5,~c^2=2.0$; (d) $a^2=1.0,~c^2=0.25$; (e) $a^2=1.0,~c^2=0.5$.}
\end{figure}

Figure~\ref{fig:orbitcompare} provides a visual comparison of a chaotic orbit starting from the same initial condition and evolving in the zeroth-order and exact potentials of Fig.~\ref{fig:potcompare1}.
Although orbits in the two potentials differ quantitatively, in many cases they are qualitatively similar in that they explore a similar volume of phase space and have similar morphology.
This pertains to the example of Fig.~\ref{fig:orbitcompare}; however, this one example does not in any way guarantee every orbit that is chaotic in the potential of the zeroth approximation is also chaotic in the exact potential.
Statistical comparisons of orbital complexities respective to the zeroth-order and exact potentials for several Case 5 configurations appear in Fig.~\ref{fig:complexitycompare}, and these show that the complexities in the two potentials can differ considerably depending on the specific parameter set under study.

\begin{figure}
\includegraphics[width=15cm]{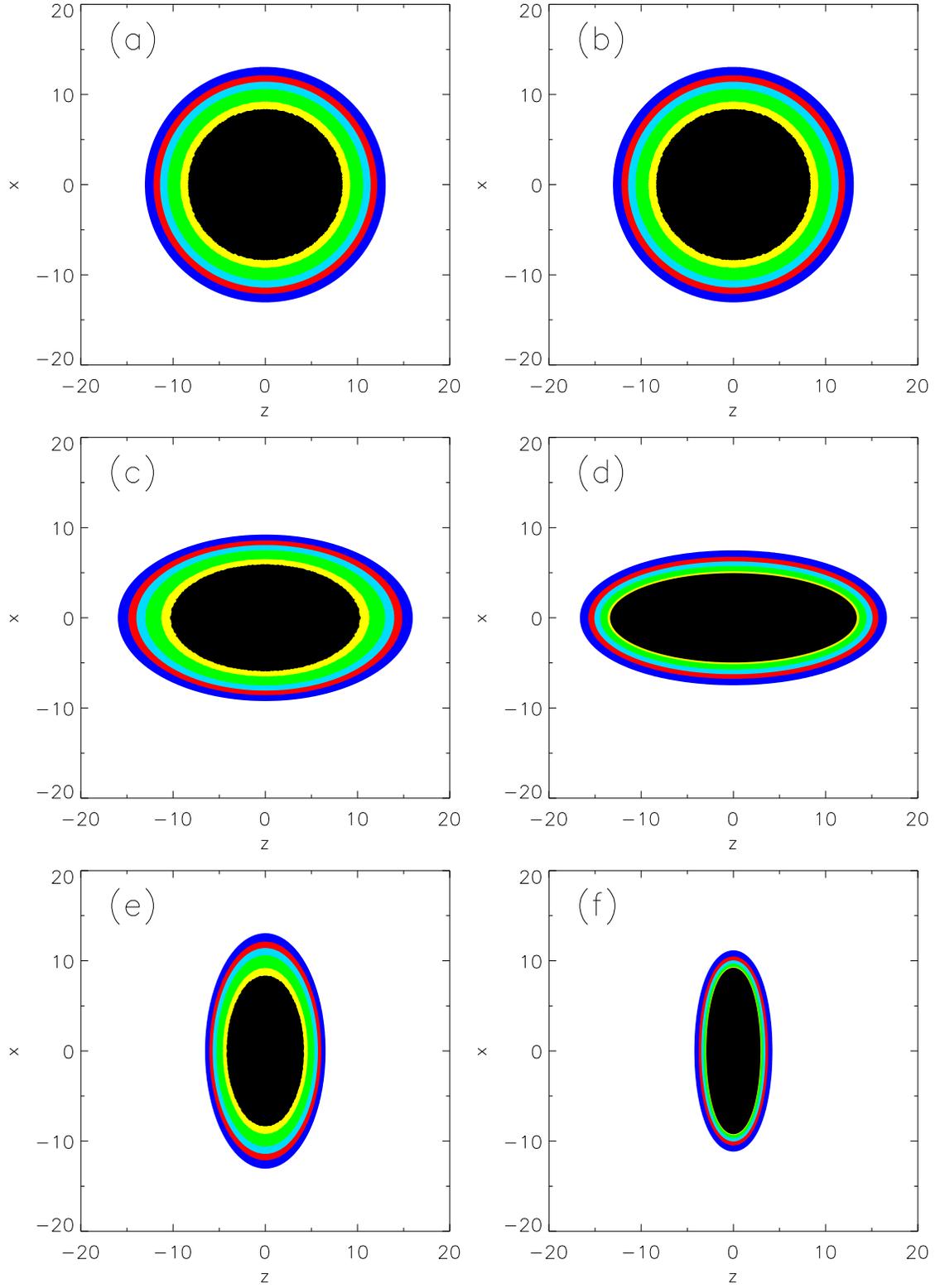}
\caption{\label{fig:densitycompare} Comparisons between the density distributions of the first approximation $n_1[R({\bf x})]$ (left) and exact solution $n({\bf x})$ (right) for Case 5 with: (a,b) $a=c=1.0$ (spherical), (c,d) $a^2=0.5,~c^2=1.5$ (triaxial), (e,f) $a^2=1.0,~c^2=0.25$ (strongly oblate spheroid).}
\end{figure}

Following the procedure delineated in Sec.~\ref{subsec:surveyresults}, we also computed the percentage of chaotic orbits in a broad range of exact Case 5 potentials.
The results, juxtaposed against their counterparts computed using $\Phi_0$, appear in Table~I and in Figs.~\ref{fig:catIa} and \ref{fig:catIb}.
For these examples, there is generally a smaller percentage of chaotic orbits in the exact potential than in the zeroth approximation.
The explanation is simple: compared to the density $n_1[R({\bf x})]$, the density derived from the exact space-charge potential $\Phi({\bf x})$ is quasi-uniform over a larger volume and falls to small values over a shorter scale length.
Accordingly, the configuration-space volume over which the space-charge force is markedly nonlinear, i.e., the Debye tail, is smaller.
Figure~\ref{fig:densitycompare} illustrates the difference in the density profiles corresponding to the approximate and exact solutions.
In the limit of spherical symmetry the profiles are identical, and they disagree more strongly as they become less spherically symmetric.
Most notable is the comparison between Figs.~\ref{fig:densitycompare}(e) and (f) concerning a strongly oblate spheroid, where we see that the corresponding exact density distribution is much more uniform than that of the zeroth approximation.
This accounts for the strong discrepancy revealed in Fig.~\ref{fig:complexitycompare}(d) concerning orbital chaoticity.
It is also consistent with expectations based on first principles: the closer a system is to being one-dimensional (e.g., sphere, cylindrically symmetric disc, infinite symmetric cylinder, in which particle motion is integrable), the less is the population of chaotic orbits.
However, the essential observation is that, in most cases, the exact TE configurations do indeed support substantial populations of chaotic orbits in keeping with expectations that surfaced from the survey based on approximate solutions.

\section{Summary, implications, and future work\label{sec:summary}}
We have explored orbital dynamics and phase mixing in thermal-equilibrium beams for which the potential is the superposition of an external potential quadratic in the coordinates and the self potential arising from space charge.
The associated parameter space spans the full range of symmetries, i.e., spherical, cylindrical, and triaxial, and the full range of density profiles, i.e., gaussian (corresponding to negligible space charge) through uniform (corresponding to maximal space charge).
To reiterate, the main findings concerning chaos in these systems, ``discovered'' in the context of zeroth approximations to the space-charge potentials and affirmed with the respective exact potentials, are:
(1) configurations corresponding to a large portion of the parameter space support considerable populations of chaotic orbits, (2) essentially all of the orbits that are chaotic reach into the Debye tail where the collective space-charge force is manifestly nonlinear, (3) prolate axisymmetric configurations support little chaos, but prolate triaxial configurations can support considerable chaos, and (4) strongly oblate spheroids support little chaos, but moderately oblate spheroids can support considerable chaos.

\begin{figure}
\includegraphics[width=15cm]{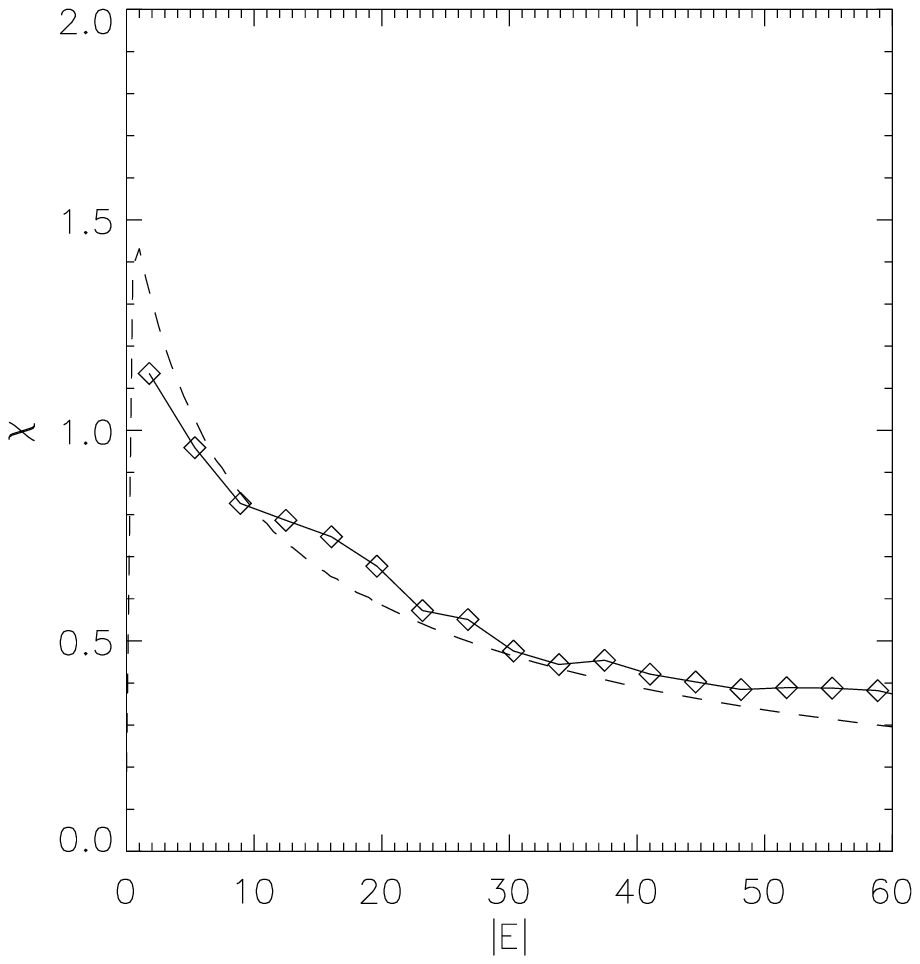}
\caption{\label{fig:mixcompare} Theoretical results (dashed curve) and numerical results (diamonds) for the mixing rate $\chi$ of chaotic orbits vs. total particle energy $|E|$ (Case 5, $a^2=4/5,~c^2=4/3$).
The unit of $\chi$ is $t_D^{-1}$.}
\end{figure}

It is of interest to compare theoretical predictions concerning TE configurations, for which we herein have established the existence of chaotic orbits, with results of our numerical experiments.
In terms of the dimensionless quantities introduced in Sec.~\ref{sec:TE}, the parameters $\kappa$ and $\rho$ of Eq.~(\ref{eq:kappasigma}) take the form
\begin{equation}
\kappa={1\over 2}\left[\Omega^2\left({1\over a^2}+1+{1\over c^2}\right)-\langle n({\bf x})\rangle\right],\;\;\;
\rho=\frac{\sqrt{\langle n^2({\bf x})-
\langle n({\bf x})\rangle^2\rangle}}{\kappa\sqrt{2}}\;.
\label{eq:kappanodim}
\end{equation}
and Eq.~(\ref{eq:analyticLyapunov}) then yields the mixing rate $\chi$.
A comparison between theory and numerical experiments appears in Fig.~\ref{fig:mixcompare} wherein the simulation results reflect statistics from initially localized clumps of 2000 particles that were started at zero velocity at various points in configuration space corresponding to various total particle energies $E$.
The figure presents a plot of the mixing rate $\chi$ versus $|E|$ in the Case 5 configuration with $a^2=4/5,~c^2=4/3$, a slightly triaxial system.
This configuration is ``not too far away'' from spherical symmetry, which means the zeroth approximation $\Phi({\bf x})=\Phi_0$ is correspondingly reasonable.
It also means only a modest population ($\sim 5\%$ for this parameter set) of chaotic orbits is supported.
The figure was derived within the framework of the zeroth approximation because therein the Lyapunov exponents, i.e., mixing rates, can be accurately computed from the simulations and the microcanonical averages required for the theory likewise can be easily and accurately evaluated.
The numerical experiments span a range $0.5\leq |E|\leq 60$, corresponding to $11\leq R\leq 25$, i.e., extending from within to well beyond the Debye drop-off in the density profile.
The agreement between theory and numerical experiments is remarkably close.

One can see from the numerical experiments described herein that chaotic mixing takes place on an e-folding time scale comparable to a dynamical time (an orbital period).
This is very fast compared to, e.g., collisional relaxation; hence, one must account for this collisionless process when designing an accelerator for the production of high-peak-current, high-brightness beams.
For example, particles comprising a beam out of equilibrium will, if globally chaotic, redistribute themselves globally and irreversibly on a dynamical time scale.
Because perturbations induced, e.g., by transitions in the beamline will drive a beam away from equilibrium, chaotic mixing can be a dynamic of practical importance.
Consider the case of a TE configuration: a small perturbation from image charges passing through an external irregularity in the beamline will distort the Debye tail.
If a substantial fraction of particles in the Debye tail are chaotic, which is the case for a wide range of bunch geometries, a corresponding fraction of the orbits comprising the distortion will quickly mix throughout the volume of the configuration.
The work done by the external perturbation in setting up the distortion will thereby appear in the form of a larger configuration-space volume.
If the perturbation is strong enough so that mixing in momentum space associated with consequent time-dependence in the potential is also substantial, then some of the work done will also appear in the form of a larger momentum space.
The net effect is a larger emittance.
If there are many such perturbations along the beamline, the cumulative emittance growth may be troublesome.

The present investigation and its associated implications concern only very specific, time-independent, single-species systems, i.e., beams (or nonneutral plasmas) in thermal equilibrium.
These are the most benign systems imaginable, yet we found even they can support chaotic orbits.
Any perturbation will create a nonequilibrium, time-dependent system that will subsequently evolve self-consistently.
Accordingly, the space-charge potential can be complicated, particularly if the perturbation is strong.
The only sensible conjecture under such conditions is that the corresponding population of chaotic orbits will be larger, and in turn chaotic mixing will be more prevalent.
Exploratory numerical simulations of an equipartitioning system and of merging beamlets have supported this notion~\cite{bohn02}.
Further exploration of time-dependent beams is warranted and will likely prove illuminating, particularly in regard to deciphering time scales for emittance growth, halo formation, etc.

By using only smooth potentials we have restricted our analysis to the six-dimensional phase space of a single particle.
Accordingly we have suppressed dissipative effects of collisions in particular, and force fluctuations in general.
Such effects can only enhance chaos, as has been demonstrated, e.g., in numerical experiments concerning self-gravitating systems~\cite{kansid02}.
As the next step, we have constructed frozen $N$-body representations of the charge densities of the TE configurations and with these representations are repeating the numerical experiments described herein.
One of our objectives is to determine the minimum number of particles needed to reproduce the dynamics associated with smooth time-independent potentials. Results will be described in a forthcoming paper~\cite{Nbody}.
In the future it will be of interest to do likewise for time-dependent systems and ultimately ascertain, e.g., conditions under which the Vlasov equation governing the six-dimensional phase space of a single particle can be applied with confidence.

\begin{acknowledgments}
We thank Henry Kandrup for helpful discussions and for financially supporting one of us (IVS) through NSF Grant AST-0070809 in the early stages of this work.
We also benefited from useful conversations with Rami Kishek.
Most of this work was supported through U.S. Department of Education Grant G1A62056.
\end{acknowledgments}


\newpage
\begin{table}[H]
Table I. Percentage of chaotic orbits: approximate vs. exact potentials\\
\begin{ruledtabular}
\begin{tabular}{cccc}
\multicolumn{4}{l}{Whole configuration space:}\\ \hline
\multicolumn{1}{c}{$a^2$}&\multicolumn{1}{c}{$c^2$}&\multicolumn{1}{c}{\% $0^{th}$}&\multicolumn{1}{c}{\% exact}\\ \hline
0.5&0.50& 6.95& 5.90\\
0.5&1.50&35.50&19.45\\
0.5&2.00&36.55&17.15\\
1.0&0.25&33.20& 7.05\\
1.0&0.50&19.85&26.05\\ \hline
\\
\multicolumn{4}{l}{Debye fall-off interval:}\\ \hline
\multicolumn{1}{c}{$a^2$}&\multicolumn{1}{c}{$c^2$}&\multicolumn{1}{c}{\% $0^{th}$}&\multicolumn{1}{c}{\% exact}\\ \hline
0.5&0.50& 7.55& 7.32\\
0.5&1.50&43.40&35.50\\
0.5&2.00&42.90&36.55\\
1.0&0.25&37.55& 7.05\\
1.0&0.50&29.45&24.55\\ \hline
\end{tabular}
\end{ruledtabular}
\end{table}

\newpage

\newpage

\newpage

\newpage

\newpage

\newpage

\newpage

\newpage

\newpage

\newpage

\newpage

\newpage

\newpage

\newpage

%\appendix
%\section{}


\begin{thebibliography}{99}
\bibitem{reiser94}M. Reiser, {\it Theory and Design of Charged Particle Beams}, (Wiley, New York, 1994), Sec. 6.2.2.
\bibitem{bohn02}C.L. Bohn, I.V. Sideris, H.E. Kandrup, and R.A. Kishek, in {\it Proceedings of the XXI Linear Accelerator Conference}, Paper TU435, http://linac2002.postech.ac.kr/, (2002).
\bibitem{kandrup94}H.E. Kandrup and M.E. Mahon, Phys. Rev. E {\bf 49}, 3735 (1994).
\bibitem{merritt96}D. Merritt and M. Valluri, Astrophys. J. {\bf 471}, 82 (1996).
\bibitem{sagan94}D. Sagan, Am. J. Phys. {\bf 62}, 450 (1994).
\bibitem{davidson99}R.C. Davidson and H. Qin, Phys. Rev. ST-AB {\bf 2}, 114401-1 (1999).
\bibitem{brown95}N. Brown and M. Reiser, Phys. Plasmas {\bf 2}, 965 (1995).
\bibitem{reiser93}M. Reiser and N. Brown, Phys. Rev. Lett. {\bf 71}, 2911 (1993).
\bibitem{reiser2}M. Reiser, {\it Theory and Design of Charged Particle Beams, op. cit.}, Sec. 5.4.3.
\bibitem{chao93}A.W. Chao, {\it Physics of Collective Beam Instabilities in High Energy Accelerators}, (Wiley, New York, 1993), p. 26.
\bibitem{lee96}For example, see {\it Space Charge Dominated Beams and Applications of High-Brightness Beams}, edited by S.Y. Lee, AIP Conf. Proc. No. 377 (AIP, New York, 1996).
\bibitem{reiser95}M. Reiser and N. Brown, Phys. Rev. Lett. {\bf 74}, 1111 (1995).
\bibitem{davidson98}R.C. Davidson, Phys. Rev. Lett. {\bf 81}, 991 (1998).
\bibitem{bohn00}C.L. Bohn, in {\it The Physics of High Brightness Beams}, edited by J. Rosenzweig and L. Serafini (World Scientific, Singapore, 2000), pp. 358-368.
\bibitem{kandrup01}H.E. Kandrup, I.V. Sideris, and C.L. Bohn, Phys. Rev. E {\bf 65}, 016214 (2001).
\bibitem{casetti96}M. Pettini, Phys. Rev. E {\bf 47}, 828 (1993); L. Casetti, C. Clementi, and M. Pettini, Phys. Rev. E {\bf 54}, 5969 (1996); L. Casetti, M. Pettini, and E.G.D. Cohen, Phys. Rep. {\bf 337}, 237 (2000).
\bibitem{goldstein50}H. Goldstein, {\em Classical Mechanics}, (Addison-Wesley, Reading, MA, 1950), pp. 228-235.
\bibitem{eisenhart29}L.P. Eisenhart, Ann. Math. {\bf 30}, 591 (1929).
\bibitem{szczesny99}J. Szczesny and T. Dobrowolski, Ann. Phys. (NY) {\bf 77}, 161 (1999).
\bibitem{cipriani98}P. Cipriani and M. Di Bari, Planet. Space Sci. {\bf 46}, 1499 (1998).
\bibitem{casetti95}L. Casetti, R. Livi, and M. Pettini, Phys. Rev. Lett. {\bf 74}, 375 (1995).
\bibitem{rosenbluth}M.N. Rosenbluth, W.M. MacDonald, and D.L. Judd, Phys. Rev. {\bf 107}, 1 (1957).
\bibitem{struckmeier96}J. Struckmeier, Phys. Rev. E {\bf 54}, 830 (1996).
\bibitem{hemsendorf02}M. Hemsendorf and D. Merritt, Astrophys. J. {\bf 580}, 606 (2002).
\bibitem{kansid01}H.E. Kandrup and I.V. Sideris, Phys. Rev. E {\bf 64}, 056209-1 (2001).
\bibitem{sideris02}I.V. Sideris and H.E. Kandrup, Phys. Rev. E {\bf 65}, 066203-1 (2002).
\bibitem{kansid02}H.E. Kandrup and I.V. Sideris, Astrophys. J. (in press).
\bibitem{Nbody}H.E. Kandrup, I.V. Sideris, and C.L. Bohn, Phys. Plasmas (in preparation).
\bibitem{kandrup02}H.E. Kandrup, in {\it Proceedings of the 2002 Athens Workshop on Galaxies and Chaos, Theory and Observations}, Springer Lecture Notes in Physics (in press).
\bibitem{bohn83}C.L. Bohn, Astrophys. J. {\bf 268}, 646 (1983).
\bibitem{efe}S. Chandrasekhar, {\it Ellipsoidal Figures of Equilibrium}, (Yale, New Haven, 1969), Theorem 12, p. 52.
\bibitem{ptvf}W.H. Press, S.A. Teukolsky, W.T. Vetterling, and B.P. Flannery, Numerical Recipes in C, (Cambridge Univ. Press, NY, 1993).
\bibitem{bennetin}G. Bennetin, L. Galgani, A. Giorgilli, and J.M. Strelcyn, Meccanica {\bf 15}, 9 (1980)  
\bibitem{complexkandrup}H. E. Kandrup, B. L. Eckstein, and B. O. Bradley, Astron. and Astrophys. {\bf 320}, 65 (1997). 
\bibitem{tabor}M. Tabor, {\it Chaos and Integrability in Nonlinear Dynamics} (Wiley, New York, 1989).
\bibitem{orbitref}R.W. Hockney and J.W. Eastwood, {\it Computer Simulation Using Particles} (IOP, London, 1988).
\bibitem{brandt}A. Brandt, Mathematics of Computation {\bf 31}, 333 (1977).
\end{thebibliography}
\end{document}